\documentclass[11pt, oneside, reqno]{amsart}   	
\usepackage{geometry}                		
\usepackage{amssymb}
\usepackage{mathrsfs}
\usepackage{latexsym}
\date{}							

\begin{document}
\begin{center}
Commutatively  Deformed General Relativity:  Foundations, Cosmology, and Experimental Tests\\ 
\vspace{2ex}
P. G. N. de Vegvar \\ 
\vspace{2ex}
SWK Research\\
1438 Chuckanut Crest Dr., Bellingham, WA 98229, USA\\
  \vspace{2ex}
Paul.deVegvar@post.harvard.edu\\ 
\vspace{2ex}
\end{center}
An integral kernel representation for the commutative $\star$-product on curved classical spacetime is introduced.
Its convergence conditions and relationship to a Drin'feld differential twist are established. A
$\star$-Einstein field equation can be obtained, provided the matter-based twist's vector generators are fixed to self-consistent values during the variation in order to
maintain $\star$-associativity. Variations not of this type are non-viable as classical field theories. $\star$-Gauge
theory on such a spacetime can be developed using $\star$-Ehresmann connections. 
While the theory preserves Lorentz invariance and background independence, the standard ADM $3+1$ decomposition of 4-diffs
in general relativity breaks down, leading
to different $\star$-constraints.  No photon or graviton ghosts are found on $\star$-Minkowski spacetime.
$\star$-Friedmann equations are derived and solved for $\star$-FLRW cosmologies. Big Bang Nucleosynthesis restricts expressions for the twist generators.
Allowed generators can be constructed which account for dark matter as arising from a twist producing non-standard model matter field.
The theory also provides a robust qualitative explanation for the matter-antimatter asymmetry of the observable Universe.
Particle exchange quantum statistics encounters \emph{thresholded} modifications due to violations
of the cluster decomposition principle on the nonlocality length scale $\sim10^{3-5} \,L_P$. Precision Hughes-Drever measurements of spacetime anisotropy
appear as the most promising experimental route to test deformed general relativity.

\vspace{2.8in}
\pagebreak

\section{Review and Introduction}
A few years ago Hopf algebra methods for commutatively deformed classical curved manifolds were studied and applied to investigate spacetime physics and
gravitation \cite{DGRv1.0}\cite{Minkowski_Procs17}. This approach differs from quantum non-commutative models, where coordinates $x^{\mu}$ are promoted to quantum operators $\hat{x}^{\mu}$ obeying
$ [ \hat{x}^{\mu}, \hat{x}^{\nu} ] = i \theta^{\mu\nu}$ for some background field $\theta^{\mu\nu}$ \cite{Dopplicher}\cite{Douglas}\cite{Chamseddine}. 
It is also distinct from non-commutatively deformed classical manifolds,
which introduce a $\star$-product of smooth functions $f, g$ with $(f\star g)(x) \ne (g \star f) (x)$ \cite{Aschieri_2}\cite{Schenkel}. Instead, commutatively deformed classical manifolds possess a commutative 
$\star$-product,  $(f\star g)(x) = (g \star f) (x)$ \cite{Lizzi}\cite{Galluccio}\cite{Ardalan}. In the early work \cite{DGRv1.0}, $f \star g$ was defined via a Drin'feld differential twist (DDT):
\begin{align} 
(f \star g)(x)  & \doteq \exp \big[ -\frac {\lambda}{2} \,\theta^{AB} \,\mathcal{L}_{X_A } \otimes \mathcal{L}_{X_B } \big] \triangleright (f \otimes g) (x) \label{DDT} \\
& = \sum _{n=0}^{\infty} \frac {(-\lambda /2)^{n} } {n!} [ \theta ^{A_{1} B_{1}} \cdots \theta ^{A_{n} B_{n}} ] [ \mathcal{L}_{X_{A_1}} \cdots  \mathcal{L}_{X_{A_n}} \, f(x) ] \\
 & \quad\quad\quad [ \mathcal{L}_{X_{B_1}} \cdots  \mathcal{L}_{X_{B_n}} \, g(x) ]. \nonumber
\end{align}
Here $\lambda$ is a dimensionful real constant, $\theta^{AB}$ is a real symmetric numerical matrix, and the $X_A$ are a set of smooth vector fields called twist generators labelled by $A$. $\mathcal{L}_v$
is the standard Lie derivative for vector field $v$, and the summation convention applies to repeated indices. To keep the $\star$-product
associative one imposes that all their Lie brackets (commutators) vanish: $[X_A , X_B ] = 0$ \cite{Schenkel}\cite{AC_Abel}. This product is also commutative, bilinear in $f$ and $g$, real $\overline{(f\star g) (x)}=
(\overline{g} \star\overline{f})(x)$, and unital $f\star 1 = f$. The $\star$-product is inserted into the action expressed in terms of field components (coefficient functions) wherever a standard product appears. 
To ensure this deformed action remains background 
independent and  Lorentz invariant, the generators must be derived from matter fields. By counting degrees of freedom one can establish that there are at most $N_X\le 2$  generators $X_A$, 
and only one comes from matter currents; the other ``twin'' generator, if it is present, being determined from the vanishing Lie bracket. For the deformed action to retain gauge invariance(s), $X$ itself must be 
gauge invariant. It was shown in earlier work \cite{DGRv1.0}
that if $X$ is derived from a matter current, then that timelike current is either a $U(1)$ current or gaugeless, and the twist producing substance is 
called Groenewold-Moyal (GM) matter. When applied to spacetime physics, the $\star$-product means the action must be invariant under $\star$-diffeomorphisms ($\star$-diffs) where the generator(s) 
control how coordinate reparametrizations (coord reparams) act on objects like $\star$-tensors. The $\star$-diffs were demonstrated to obey a different Lie algebra than standard diffs, leading to distinct
spacetime physics called deformed general relativity (DGR). The $\star$-product is nonlocal over an $X$ dependent proper length $\xi_c$. An intuitive way to picture $\star$-deformed spacetime
is that events are no longer pointlike, but instead become smeared over the length scale $\xi_c$ on an $N_X$-dimensional submanifold spanned by the smooth
vector generators $X_A$ passing through the event. DGR is still a metric theory of gravity. However while $X$ is constructed from GM matter, it also plays a non-standard role in determining geometry, 
aside from GM matter fields entering the stress-energy tensor in the usual way.
Constructing $X$ from GM matter as scalar or fermionic currents and using dimensional analysis was shown to imply the existence of a window for $\lambda$
and GM particle mass $m_{\mathrm{GM}} \sim 1$ TeV that can produce $\xi_c /L_P \simeq 10^{3-5}$, where $L_P \simeq 1.6\cdot 10^{-35}$ m is the Planck length. The GM gauge current expressions for 
the twist generator read 
\begin{align}
X^{\mu}(\phi) & = \left\langle \binom{\mathrm{Re}} {\mathrm{Im}} \sum_{l} Q_{l}\, g^{\mu\nu}\star \phi_{l}\star ( D_{\nu} \phi)_{l}^{\dagger}\right\rangle\quad\text{for GM scalar } \phi ,\text{and} \label{XDPhiN} \\
X^{\mu}(\psi) & = \left\langle  \binom{\mathrm{Re}} {\mathrm{Im}} \left(\sum_{l} Q_{l}\, \bar{\psi}^{l}\star \gamma^{I}e^{\mu}_{I}\star\psi^{l}\right) \right\rangle\quad \text{for GM Dirac fermion } \psi  , \label{XFCN}
\end{align}
where $l$ is a GM species index, $Q_l$ is the $U(1)_{GM}$ charge or some other quantum number, $e^{\mu}_{I}$ is the $\star$-tetrad, and $g^{\mu\nu}$ are components of the $\star$-inverse metric tensor.
$(D_{\mu}\phi)_{l}\doteq \partial_{\mu}\phi_{l}-i\,A_{\mu} ^{\alpha} \,(\hat{t}_{\alpha})^{m}_{l}\star\phi_{m}$ is the $\star$-gauge covariant derivative, with the $\hat{t}_{\alpha}$ being the Lie (matrix) 
generators for all the gauge interactions of $\phi$, $\alpha$ labeling which gauge generator. The angle brackets denote an expectation value
with respect to some GM state or particle ensemble described by a density operator $\hat{\rho}$ as $\langle \cdots \rangle \doteq \mathrm{Tr}\;(\hat{\rho} \cdots),$ and the brackets render $X$ a \emph{classical} vector field.
These expressions for the twist generator themselves
contain the $\star$-product and not the standard one because the DDT (\ref{DDT}) will insert arbitrary powers of $X$ into the action, and for the action to be $\star$-diff invariant means a composite object like $X$
must incorporate $\star$. Because $\star$ contains $X$ and $X$ contains $\star$, $X$ is defined recursively or self-consistently (sc)  as a solution of $X=X_{\mathrm{sc}}[\phi , *(X)]$, 
where $X_{\mathrm{sc}}$ is an expression such as (\ref{XDPhiN}) or (\ref{XFCN}), and will be discussed in more detail below. 
Investigating these GM gauge current forms for the generators in \cite{DGRv1.0}
led to the conclusion that GM matter
 could not be standard model (SM) matter, rather it had to  be either a sterile neutrino or a viable dark matter (DM) candidate. 
 A $\star$-Einstein field equation for the $\star$-metric tensor was derived that included an uncalculated $\mathscr{O}(
 \lambda ^1)$ correction term $\Delta_{\mu\nu}$ to the $\star$-Einstein tensor coming from variations of the fields contained within $\star(X)$.
  Estimates for the experimental or observational deviations from classical GR and the standard model of particle physics led to 
 vanishingly small effects primarily owing to the near Planckian scale of $\xi_c$. Classical DGR is anticipated to hold only up to energy scales $\mu\lesssim \xi_c^{-1}$ in natural units $\hbar = c=1$,
 since at greater energies quantum corrections (fluctuations) of $X$ are expected to disrupt the classical picture.\\
 
 These early steps raise questions about the foundations of DGR along with its cosmological and phenomenological implications. For instance, when using the the DDT  form (\ref{DDT}) for the twist
 the $\star$-Einstein equation, and indeed any $\star$-Euler-Lagrange or $\star$-wave equation, all become infinite order partial differential equations.  The theory then appears to lose predictive 
 power, since knowledge of derivatives
 of all orders at some boundary condition  becomes necessary to determine the dynamics. On the other hand, truncating the DDT twist to some finite order in $\lambda$ would produce a PDE of finite order, but
  this PDE is necessarily local and cannot capture DGR's essential nonlocality. Hence it would be desirable to find an integral form for the $\star$-product in curved spacetime, 
  which might also be more suitable for quantization of DGR. Additionally, one would like to calculate
  the $\Delta_{\mu\nu}$ correction term in the $\star$-Einstein equation explicitly and apply the result to cosmology, such as $\star$-FLRW models. It could also prove fruitful to study the GM gauge current
  expressions (\ref{XDPhiN}) and (\ref{XFCN}) for the generator in greater detail and to seek experimental tests for DGR. The foundations of DGR deserve further scrutiny as well: 
  What are $\star$-gauge theories like, and how do they differ from standard gauge theory? What about DGR's canonical structure and the suitability of loop quantization methods for its quantum limit?\\
  
 The following sequential outline lists the topics to be discussed:\\
 
 Section 2: $\star$-differential geometry,  non-standard push-forward ($\star$-coord reparams), Nice Basis \\
 
 Section 3: Integral kernel twist (IKT),  convergence conditions, relationship to the DDT \\
 
 Section 4: Simple dynamics and the $\star$-Einstein equation  \\
 
 Section 5: $\star$-gauge theory: $\star$-fiber bundles, $\star$-Bianchi identity, $\star$-holonomies \\
 
 Section 6: Photon and graviton $\star$-ghosts banished from Minkowski spacetime \\
 
 Section 7: Canonical structure of DGR: No ADM (3+1) decomposition of $\star$-4-diffs, classical non-viability of $\delta X\ne0$ variations, the self-consistency process (deformation flow) \\
 
 Section 8: $\star$-FLRW cosmologies, Big Bang Nucleosynthesis constraints on $X$ and new generators that satisfy them,  GM as DM or SM, $\star$-scalar wave equations, 
antimatter in DGR, no spectral dimension reduction in classical DGR \\
 
 Section 9: Experimental Tests of DGR: Violations of the cluster decomposition principle, spin-statistics and CPT theorems, and Pauli Exclusion Principle; Michelson-Morley and
  Hughes-Drever measurements \\
 
 Section 10: Conclusions and perspective \\ 
  
 \section{$\star$-differential geometry and $\star$-coord reparams}
 
 Some basic aspects of $\star$-differential geometry are presented here that are useful for both deeper understanding as well as for practical calculations. It is assumed that $\star$ is associative, commutative,
 real, and unital. \\
 
 The first noteworthy point is if $f$ is a real or complex-valued smooth function on some manifold $M$ with a $\star$-product (a $\star$-manifold), $\omega = \mathrm{d} f$  a $\star$-1-form field, 
 and $V$ a $\star$-vector field on $M$, then their $\star$-inner product ($\star$-contraction) obeys
 \begin{align}
 \langle V, \mathrm{d} f\rangle_{\star} & = \langle \omega , V\rangle_{\star} = V^{\mu} \star \omega_{\mu} \\
& \ne V^{\mu} \cdot \omega_{\mu}  = \langle V, \mathrm{d} f \rangle_{\cdot}  = V^{\mu} \cdot (\partial_{\mu} f) \doteq V[f]\doteq \mathcal{L}_{V}\, f , \label{CContraction}
\end{align} 
where $ \mathcal{L}_{V} f$ is the standard Lie derivative of $f$ with respect to $V,$ and the exterior derivative $\mathrm{d}$ is undeformed. This affects how $\star$-push forwards operate
between the tangent spaces of $\star$-manifolds \cite{Nakahara_pushfwd}.
 Let $f:M\to N$ be a smooth map between $\star$-manifolds $M$ and $N,$ and take $p\in M.$ Let $V=V^{\mu} \, (\partial /\partial x^{\mu}),$ 
where $x^{\mu} = \varphi ^{\mu}(p)$ are coordinates into $M$, and we seek the $\star$-push forward of $V$: $W^{*} = f^{(\star)}_{*} \, V = W^{\alpha} \, (\partial/\partial y^{\alpha}),$ where 
$y^{\alpha} = \psi ^{\alpha} (f(p))$ are coordinates into $N.$ The standard (dot product) push forward acts as $f_{*}:V\in T_{p}M \mapsto W=f_{*} \, V \in T_{f(p)} N.$ The standard dot product definition
introduces a smooth function $g: N\to \mathbb{R},$ and then defines $(f_{*} \, V)[g] \doteq V[g \circ f] =\langle \mathrm{d}(g\circ f), V\rangle_{\cdot}^{M}.$ 
Here the superscripts $M,N$ designate on which manifold the contraction is to be taken.
However by (\ref{CContraction})
the last equality does not apply for the $\star$ case. So by choosing the $\star$-push forward to be defined via \emph{differentials} instead of Lie derivatives, one sets
\begin{equation}
(f^{(\star)}_{*}\, V)[g] \doteq \langle \mathrm{d} (g\circ f ), V\rangle^{M}_{\star} = \big(\partial_{\mu} (g\circ f)\star V^{\mu}\big)^{M}.
\end{equation}
Now setting $g = y^{\alpha}$ or $\psi^{\alpha}$ yields for the components
\begin{equation}
(W^{*})^{\alpha} = (f^{(\star)}_{*} V)^{\alpha} = \Big( V^{\mu} \star \frac {\partial y^{\alpha}} {\partial x^{\mu}}\Big)^{M},\label{Vec_xform}
\end{equation}
the transformation rule of $\star$-vectors under a $\star$-coord reparam ($\star$-diff). This is readily extended to type $(q,0)$ (contravariant) $\star$-tensors. It also straightforward to apply
similar reasoning to $\star$-pullbacks  acting on forms (but in the opposite direction): $f^{(\star)*}: T^{*}_{f(p)}(N) \to T^{*}_{p} (M)$ by duality. Take $\omega\in T^{*} N,$ a $\star$-one form 
field on $N$, and define
\begin{equation} 
\langle f^{(\star)*} \omega, V\rangle_{\star}^{M} \doteq \langle \omega, f^{(\star)}_{*} V \rangle_{\star} ^{N}. 
\end{equation}
In terms of components, $\omega = \omega_{\alpha} \,\mathrm{d}y^{\alpha} \in T^{*}N$, $f^{(\star)*} \omega = \xi_{\mu} \, \mathrm{d} x^{\mu} \in T^{*}M$ this reads
\begin{equation}
\xi_{\mu} = \Big( \omega_{\alpha} \star \frac {\partial y^{\alpha}} {\partial x^{\mu}} \Big)^{N}. \label{Form_xform}
\end{equation}
Eqns. (\ref{Vec_xform}) and (\ref{Form_xform}) clearly generalize to $(q,r)$ type $\star$-tensors. One concludes that on a single $\star$-manifold the $\star$-transformation rule is simply the well-known
standard product form with $\star$ replacing $\cdot$ throughout. This provides a concrete representation for $\star$-coord reparam covariance. \\

What about the chain-rule for such $\star$-transformations? Requiring (\ref{Vec_xform}) and (\ref{Form_xform}) to be consistent across
two successive maps taking manifolds $K\to M\to N$ compared to directly as $K\to N$ leads to the following.
Suppose $K,M,N $ are $\star$-manifolds linked by $\star$-diffs $g:K\to M, f:M\to N$ choose a point $a\in K$ and define the mapping between tangent spaces 
$D_{p}^{(\star)}(f)\doteq  f^{(\star)}_{*}(p)$ for $p\in M$, then
\begin{equation}
D^{(\star)}_{a}(f\circ g) = \left( D^{(\star)}_{g(a)} (f) \right) \circ \left( D^{(\star)}_{a}(g) \right),
\end {equation}
with $\star$-Jacobian matrices $J^{(\star)}$ associated with $D^{(\star)}$ acting as
\begin{equation}
J^{(\star)}_{f\circ g} (a) = J^{(\star)}_{f}(g(a)) \star_{M} J^{(\star)}_{g}(a).
\end{equation}
That is, the ordinary chain-rule simply acquires a suitable $\star$ to become a $\star$-chain-rule. \\

It is also useful to recall the concept of a ``Nice Basis'' (NB) for a special class of twists \cite{Aschieri_2}\cite{Schenkel}. 
Suppose one has a set of $N_X$ vector field twist generators 
$X_A$ for the DDT on a $\star$-manifold $M$ that are pointwise linearly independent and obey $[X_{A}, X_{B}]=0$ for all $A,B$ (a so-called Abelian twist). Then there is a set of smooth basis vector fields 
$\{ e_a \},$ $a=1,\ldots , \mathrm{dim} \, M$, such that $[e_{a}, e_{b}]=0$ for all $a,b$ and all the $X_A$ commute with all the $e_b , [X_{A}, e_{b}]=0,$ for all $A,b$ at all points of $M$. This choice of basis
is called a Nice Basis. As discussed in Schenkel using this NB one can introduce $\star$-Levi-Civita connections as
\begin{equation}
\Gamma^{\star\lambda}_{\mu\nu} = \left( \frac{1}{2} \right) \, g^{\lambda\kappa} \star \left[ e_{\mu}(g_{\nu\kappa}) + e_{\nu} (g_{\mu\kappa})-e_{\kappa}(g_{\mu\nu} )\right] ,
\end{equation}
where $e_{\mu}(g_{\nu\kappa}) \doteq \mathcal{L}_{e_{\mu}}\, g_{\nu\kappa}$ is the standard Lie derivative of the
\emph{component} (coefficient \emph{function}) $g_{\nu\kappa}.$ The corresponding $\star$-metric compatible, $\star$-torsion-free $\star$-covariant derivative is denoted $\stackrel {\star} {,}$ to distinguish it from
the standard covariant derivative $;$ notation.  In  component form it reads 
\begin{equation}
T^{\mu_1 \cdots \mu_q}_{\nu_1 \cdots \nu_r \stackrel{\star}{,} \lambda} =  e_{\lambda} (T^{\mu_1 \cdots \mu_q}_{\nu_1 \cdots \nu_r} ) - T^{\mu_1 \cdots \mu_q}_{\sigma \cdots \nu_r} \star
 \Gamma^{\star\sigma} _{\lambda\nu_1} - \cdots 
  + T^{\mu_1 \cdots \sigma}_{\nu_1 \cdots \nu_r} \star \Gamma^{\star\mu_q}_{\lambda\sigma}.
 \end{equation}
A Nice Basis deserves its name because usually the Lie derivative does not obey the $\star$ -Leibniz rule:
\begin{equation}
\left[ \mathcal{L}_{v} (f\star g) \right] (x) \ne \left[ (\mathcal{L} _{v}\, f) \star g + f \star (\mathcal{L}_{v} \, g) \right] (x) \label{L-non_Leib}
\end{equation}
for functions  $f,g$ and general vector field $v$. However, if $v$ is any of the NB vector fields $e_a$, then
\begin{equation}
\left[ \mathcal{L}_{e_a} (f\star g) \right] (x) = \left[ (\mathcal{L} _{e_a}\, f) \star g + f \star (\mathcal{L}_{e_a} \, g) \right] (x),
\end{equation} 
and the Lie derivatives with respect to any nice basis vector field  \emph{are} $\star$-Leibniz. One may anticipate that the NB will be a natural basis choice for calculations.
For example in such a basis the $\star$-covariant derivative is also $\star$-Leibniz, so $\star$-contraction and $\star$-covariant differentiation commute there, transforming the Nice Basis
into a veritable Land of Oz for $\star$-tensor analysis.\\

For instance, the standard derivation \cite{WGC_p376} of the Killing vector relation $\xi_{\sigma ; \rho} +\xi_{\rho ; \sigma} = 0$ does not generally apply to $\star$-manifolds 
unless one restricts oneself to the NB because 
argument uses the Leibniz rule. Adopting the convention of denoting components in a NB by carets on indices, one does have 
\begin{equation}
\xi_{\hat{\sigma} \stackrel {\star} {,} \hat{\rho}} +\xi_{\hat{\rho} \stackrel {\star} {,} \hat{\sigma}} = 0 \label{NB_Killing}
\end{equation}
in a NB. Since this is the vanishing of
a $\star$-tensor (i.e., the LHS is $\star$-coord reparam covariant), it is a $\star$-coord reparam invariant statement, and so the $\star$-version is true in any basis as well.
 Meaning one can remove the carets in eqn. (\ref{NB_Killing}) with impunity. \\
 
 It is also important to realize the exterior derivative $\mathrm{d}$  remains \emph{undeformed;} i.e., for $f,g \in C^{\infty}(M),$
 \begin{equation}
 \mathrm{d} (f\star g) = (\mathrm{d} f) \star g + f\star (\mathrm{d} g), \label{d_undefd}
 \end{equation}
 that is, $\mathrm{d}$ remains $\star$-Leibniz \cite{Schenkel}. This turns to be important for $\star$-gauge theory.
 How can eqn. (\ref{d_undefd}) be compatible with the non-Leibniz property of the Lie derivative (\ref{L-non_Leib})?
 Noting $\mathrm{d} f = (\partial_{\mu} f)\cdot \theta^{\mu}$ where $\theta^{\mu}\doteq \mathrm{d} x^{\mu}$ are one-forms, then
 \begin{align}
 \mathrm{d} (f\star g) & = \theta^{\mu} \cdot\big[ \partial_{\mu} (f\star g) \big]  \nonumber  \\
 & = (\mathrm{d} f) \star g + f\star (\mathrm{d} g) = (\theta^{\mu} \cdot \partial_{\mu} f) \star g + f \star (\theta^{\mu} \cdot \partial_{\mu} g) \nonumber \label{NAL1}. \\
 & \ne \theta^{\mu} \cdot \big[ (\partial_{\mu} f)\star g +f \star(\partial_{\mu} g) \big],
 \end{align}
 because $f \cdot (g\star h) \ne (f \cdot g)\star g$ due to the mutual non-associativity of the dot and star products.
 Then one obtains
 \begin{equation}
 \theta^{\mu} \cdot \big[ \partial_{\mu} (f\star g) - (\partial_{\mu} f) \star g - f\star (\partial_{\mu} g) \big] \ne 0,
 \end{equation}
 so one \emph{cannot} conclude that
  \begin{equation}
 \partial_{\mu}(f\star g) = (\partial_{\mu} f)\star g + f\star (\partial_{\mu} g). \quad\text {(FALSE)}
\end{equation}
The fact that the \emph{exterior} derivative is $\star$-Leibniz does \emph{not} imply the \emph{Lie} derivative is also $\star$-Leibniz; or equivalently
non-$\star$-Leibniz Lie derivative does not imply non-$\star$-Leibniz exterior derivative. \\

\section{The integral kernel twist}

An integral form of the twist on \emph{flat}  (Euclidean or Minkowskian) spacetimes was first introduced by Galluccio and coworkers in 2009 \cite{Lizzi}\cite{Galluccio}. 
We  briefly recapitulate that Fourier-based approach as it motivates
the integral twist on curved manifolds, even though because of its restriction to flat manifolds it is inapplicable to gravitational physics. 
Start from the familiar Fourier transform of a smooth function $f(x)$ on a $d$-dimensional flat manifold $\mathcal{M}$,
\begin {equation}
\tilde{f} (q) \doteq (2\pi)^{-d/2} \int_{M} \mathrm{d}^{d}x \, \exp [-iq\cdot x] \,f(x). \label{Std_FFT}
\end{equation}
Then the Galluccio $\star$-product is defined as
\begin{equation}
(f \star_{\mathscr{G}} g)\, (x) \doteq (2\pi)^{-d} \int \mathrm{d}^{d} p\, \mathrm{d}^{d} q\, \tilde{f}(q)\, \tilde{g}(p-q) \exp \left[ ipx+\tilde{\alpha}(p,q) \right], \label{Galluccio_IKT}
\end{equation}
where the desired properties of the $\star$-product constrain $\tilde{\alpha}$ as: \\

(1) Associativity: $\tilde{\alpha}(p,q) + \tilde{\alpha}(q,r) = \tilde{\alpha}(p,r) + \tilde{\alpha}(p-r,q-r)$ \\

(2) Commutativity: $\tilde{\alpha}(p,q) = \tilde{\alpha}(p,p-q)$\\

(3) Unitality: $\tilde{\alpha}(p,p) = 0 = \tilde{\alpha}(p,0)$, and \\

(4) Reality: $\overline{\tilde{\alpha}(p,q) }= \tilde{\alpha}(-p,q-p).$ \\

Setting $\tilde{\alpha}(p,q) = \beta(q) - \beta(p)+\beta(p-q)$ assures both associativity and commutativity. The function $\beta(q)$ is called a 2-cocycle.
 Reality then constrains $\overline{\beta(q)} = \beta(-q),$ so the real (imaginary) part of $\beta(q)$ is $q$-even (odd).
Finally, unitality is simply $\beta(0) = 0.$  To be equivalent to a DDT means $\beta$ and $\tilde\alpha$ are analytic functions of their arguments. 
The standard product is just $\beta(q)=0$ for all $q.$ The next order na\"ive guess $\beta(q) = 
(k\cdot q)^2$ for some constant $d$-vector $k$ is a DDT,  but it violates  strict convergence conditions, so that the Fourier transforms $\tilde{f}(q)$ and $\tilde{g}(q)$ must have 
Gaussian roll-offs at large $\| q\|.$ This occurs because for this choice of $\beta,$ $\alpha(q_1 ,q_2 ) \doteq \tilde{\alpha} (q_1 +q_2, q_1 )  = (- 2) (k\cdot q_1)(k\cdot q_2)$ 
does not have uniform sign on $(q_1 ,q_2)$-space. 
To avoid having to place restrictions on the spectra of $f,g$, it would be desirable to have $\alpha(q_1, q_2)$ be bounded above by some real constant. This will be examined further below.
\\

The Galluccio $\star$-product's use of Fourier methods is by no means accidental, it encodes the translational invariance of the flat geometry it inhabits. Indeed, the Fourier kernel $\exp [-i q\cdot x]$
is the eigenfunction of the translation operator. The intuition behind the Galluccio twist is that $\exp \tilde{\alpha}(p,q)$ filters momenta in the standard convolution of Fourier transforms 
corresponding to the Fourier transform of the dot-product. Harmonic analysis on a group requires that group to be locally compact and Abelian, here that would be the group of translations on flat $d$-space(time).
A general curved manifold, however, does not possess such a high degree of symmetry, initially suggesting that any version of the Galluccio twist would be doomed in a curved habitat. 
An integral kernel twist (IKT) for
gravitational physics calls for some other insight. And here it is: Notice that classical $\star$-manifolds come equipped with a set of $N_X \le 2$ smooth vector field twist generators $X_A$. 
As a result one can construct a set of one-parameter 
locally compact Abelian groups from those vector fields by their integral curves. More specifically, choose any point $x$ on $\mathcal{M}$ (now generally curved) 
with coordinates $x^{\mu}$, let $Y=X_A$ for some $A\in \{1, \ldots ,N_{X} \},$ and define for $t\in \mathbb{R}$
\begin{equation}
\frac {\mathrm{d} \sigma ^{\mu}_{Y}(x, t)} {\mathrm{d}t}  \doteq Y^{\mu} \big(\sigma_{Y}(x,t) \big) , \label{Congruence_ODE} 
\end{equation}
\begin{equation}
\sigma^{\mu}_{Y}(x,0)  \doteq x^{\mu}. \label{Congruence_IC}
\end{equation}
$\sigma^{\mu}_{Y} (x,t)$ is the congruence through $x$ generated by $Y$ and parametrized by $t.$
Since $Y$ is any of the $N_X\le 2$ vector field twist generators $X_A$, there is a family of $N_X$ such congruences. Moreover, the generators satisfy $[X_{A}, X_{B} ] =0$ on all of $\mathcal{M}$ for all
$A,B$. Recalling that the Lie-commutator of two vector fields is the ``closer of quadrilaterals'' \cite{Nakahara_Commtr}, the $N_X$ $\sigma$-displacements from any $x\in \mathcal{M}$ 
generated by (\ref{Congruence_ODE})
can be taken in any order or partitioned in any way. So given any such $x\in \mathcal{M}$ one can parametrize a $N_X$-dimensional smooth submanifold of $\mathcal{M}$ containing $x$ 
by the corresponding real-valued
$N_X$-tuples $(t_{1}, \ldots ,t_{N_X})$ and there is a unique point in this submanifold for each distinct $N_X$-tuple of $t$'s and vice-versa. Call the submanifold containing $x$ the ``twist submanifold of $x$,''
abbreviated as TSM$(x)$. Clearly any two TSMs are either identical or disjoint, and the TSMs partition $\mathcal{M}$ (or that part of $\mathcal{M}$ where at least one generator does not vanish so 
there is a nontrivial $\star$-product there). \\

Now, what about the Fourier transforms? Given any $x\in \mathcal{M}$ one may define the Fourier-like transform based at $x$ of a ($\mathbb{C}$-valued) function $f$  on TSM$(x)$ by
\begin{equation}
\tilde{f} _{x} (\vec{q}) \doteq (2\pi)^{-N_X} \int _{\mathbb{R}^{N_X}}  \mathrm{d}^{N_X} t \, \exp[-i \vec{q} \cdot \vec{t} ]\, f[\sigma(x,\vec{t})], \label{def_FFT}
\end{equation}
where $\vec{t}, \vec{q}$ are $N_X$-tuples and the dot product in the argument of $\exp$ is the standard Euclidean one. Loosely speaking, $\vec{q}$ is an $x$-based momentum conjugate 
to the $\sigma$-displacement starting at $x$ parametrized by $\vec{t}$. Essentially this is just harmonic analysis on the TSMs, which have been equipped with a locally compact Abelian group
based on twist generators.
In undeformed but curved spacetime there are no nontrivial (non-zero) and 
commuting $X_A$ to effect such an harmonic analysis. \\

Onto the IKT:  One constructs the curved IKT for scalar (coefficient) functions $f,g$ by replacing $\tilde{f}(q) $ from eqn (\ref{Std_FFT}) in the flat spacetime Galluccio $\star$-product 
eqn. (\ref{Galluccio_IKT}) by $\tilde{f}_{x}(q) $ from eqn. (\ref{def_FFT})  and similarly for the transform of $g$. Collecting factors, the result reads
\begin{align}
(f\star g)(x)  \doteq &\; (2\pi)^{-2N_X}\,\int  \mathrm{d}^{N_X} q_{1}\, \mathrm{d}^{N_X} q_{2} \,\mathrm{d}^{N_X} t_{1} \, \mathrm{d}^{N_X} t_{2} \;\exp \big[-i \vec{q}_{1} \cdot \vec{t}_{1} - i \vec{q}_{2} \cdot \vec{t}_{2}\big] 
\label{New_STAR} \\
& \times\exp \big[\alpha(\vec{q}_{1}, \vec{q}_{2}) \big] \; f\big[\sigma(x, \vec{t}_{1})\big] \; g\big[\sigma(x, \vec{t}_{2})\big],\quad \text{where} \nonumber \\
\alpha(\vec{q}_{1}, \vec{q}_{2} )  \doteq &\; \tilde{\alpha} (\vec{q}_{1}+\vec{q}_{2}, \vec{q}_{1})  = \beta(\vec{q}_{1}) + \beta(\vec{q}_{2}) - \beta(\vec{q}_{1} + \vec{q}_{2}). \label{New_Alpha}
\end{align} 
Provided the $X_A$ themselves are $\star$-coord reparam covariant smooth $\star$-vector fields, eqn. (\ref{Congruence_ODE}) is equivalent to the vanishing of a $\star$-vector field, and consequently definition 
(\ref{New_STAR}) is also $\star$-coord reparam invariant, so the $\star$-product of two $\star$-scalar functions is again a $\star$-scalar.\\

It is now straightforward to verify that the IKT eqn. (\ref{New_STAR}) has all the requisite desiderata. Beginning with unitality, one has (dropping now superfluous $N_X$-superscripts and vector accents)
\begin{align}
(f\star 1)(x) & = (2\pi)^{-N_X} \,\int \mathrm{d}q_{1}\,\mathrm{d}q_{2}\,\mathrm{d}t_{1}\,  f[\sigma(x,t_1) ]\, \delta^{(N_X)} (q_2) \,\exp [-i q_{1}\cdot t_{1}+\alpha(q_{1},q_{2}) ]\\
& = (2\pi)^{-N_X} \,\int \mathrm{d}q_{1}\,\mathrm{d}t_{1}\,f[\sigma(x,t_1) ]\,\exp [-i q_{1}\cdot t_{1}+\alpha(q_{1},0) ]\\
& = \int \mathrm{d}t_{1}\,f[\sigma(x,t_1) ]\, \delta^{(N_X)} (t_1)\\
& = f(x),
\end{align}
where $\alpha(q,0) = \beta(0) =0 $ was used to go from the second to third lines.\\

To verify reality, consider
\begin{align}
\overline{(f\star g)(x)} & = (2\pi)^{-N_X} \,\int \mathrm{d}q_{1}\,\mathrm{d}q_{2}\,\mathrm{d}t_{1}\,\mathrm{d}t_{2}\,\overline{f[\sigma(x,t_1 ) ]}\;\overline{g[\sigma(x,t_2 ) ]} \\
& \quad\times\exp [ iq_{1}\cdot t_{1} + iq_{2}\cdot t_{2} +\overline{\alpha(q_1 , q_2 )}] \\
& = (2\pi)^{-N_X} \,\int \mathrm{d}q_{1}\,\mathrm{d}q_{2}\,\mathrm{d}t_{1}\,\mathrm{d}t_{2}\, \overline{f[\sigma(x,t_2 )]}\;\overline{g[\sigma(x,t_1 )]}\\
& \quad\times\exp [ -iq_{1}\cdot t_{1} -iq_{2}\cdot t_{2} +\overline{\alpha(-q_2 , -q_1 )}] \\
 & = (\overline{g} \star \overline{f})(x),
 \end{align}
  where going from the first to second line all integration variables' subscripts 1 and 2 were swapped, and then $q_j \mapsto -q_j $. To get the last line note that
  \begin{align}
  \overline{\alpha(-q_{2} , -q_{1} )} & =  \overline{\beta(-q_{2 })} + \overline{\beta(-q_{1})} -\overline{\beta(-q_{1} - q_{2 })}\\
  & =\beta(q_{2}) +\beta(q_{1})-\beta(q_{1}+ q_{2}) \\
  & = \alpha(q_{1} , q_{2} ),
  \end{align}
  since $\overline{\beta(-q)}=\beta(q)$. One can go further. If $\mathrm{Im}\, \alpha \ne 0,$ then $\mathrm{Im}\, \beta(q)$ would be odd in $q$. If one expects the $\star$-product to respect time and spatial
  inversion symmetries, then it suffices to take $\mathrm{Im}\,\beta =0=\mathrm{Im}\,\alpha,$ and so $\beta(-q) = \beta(q).$  \\
 
  $\star$-commutativity readily follows from $\alpha(q_2 ,q_1) = \beta(q_1 ) + \beta(q_2) - \beta(q_1 + q_2) = \alpha(q_1 ,q_2 )$. Verifying associativity requires more algebra, but making use of
  $\sigma(\sigma(x,t), t') = \sigma(x,t+t')$ means associativity requires $\alpha(p-q,q)=\alpha(q,p-q)$, and both sides of the latter relation are found to be $\beta(p-q)+\beta(q)-\beta(p).$ The $t$-additive 
  (Abelian) property of displacements $\sigma$ plays a crucial role here. This is another example of the simplifying Abelian group structure in harmonic analysis.  
  Now that one has a curved IKT, one may forget its flat space Galluccio analog. However it would have been difficult to jump directly to the IKT representation eqn. (\ref{New_STAR}) 
  without using the Galluccio twist as a springboard. \\
  
  What can one say about the convergence properties of the IKT? If $\alpha(q_1 ,q_2 ) \le B$ for some real number $B$, then 
  \begin{align}
  \big| (f\star g)(x) \big| & \le \int \mathrm{d} q_{1}\, \mathrm{d} q_{2}\, \exp \big[\alpha(q_1, q_2 ) \big] \,\big| \tilde{f}_{x}(q_1 )\big| \, \big| \tilde{g}_{x}(q_2 )\big| \nonumber \label{H}\\
  & \le \exp (B) \, \Bigg[ \int  \mathrm{d} q_{1}\,\big| \tilde{f}_{x}(q_1 )\big| \Bigg] \,\Bigg[ \int  \mathrm{d} q_{2}\,\big| \tilde{g}_{x}(q_2 )\big| \Bigg]. 
  \end{align}
  The rightmost quantity will be bounded provided $\tilde{f}_{x}(q )$ and $ \tilde{g}_{x}(q)$ are both $L^1$ on $q$-space, which is more general than a Gaussian drop-off.
  Also by noting for $z$ a non-zero real,  taking $f\in C^{\infty}_{0}(\mathcal{M})$  (smooth and of compact support),
 and integrating
 $ | \int \mathrm{d}x \,f(x) \exp[\pm \,ixz]\, | $ by parts $n$ times to get 
 \begin{equation}
 \Bigg|\int \mathrm{d}x \,f^{(n)}(x)\,(\pm 1/iz)^{n} \,\exp [\pm ixz] \Bigg |\le |z|^{-n} \int \mathrm{d}x \,\big| f^{(n)}(x)\big|, \\
 \end{equation}
 and then the last factor is bounded, shows 
$\tilde{f}_{x}(q )$ falls off faster than any polynomial at large $ \|q \|$ for such $f$. So  eqn. (\ref{H}) gives $|(f\star g)(x)| < \infty$
for $f, g \in C^{\infty}_{0}(\mathcal{M}).$  Alternatively, by H\"ormander \cite{Horm_Thm 7.1.5_p160}, if $f$ and $g$ lie in the Schwartz space $\mathcal{S}$ of smooth functions of rapid drop-off
\begin{equation}
\mathcal{S} \doteq \big\{ \phi \in C^{\infty}(\mathcal{M}) \big| \sup_{x} | x^{\alpha}\,\partial^{\beta}\, \phi | < \infty \quad \forall\, \alpha, \beta \big\} \\
\end{equation}
where $\alpha,\beta$ are multi-indices, then $\tilde{f}_{x} (q)$ and $\tilde{g}_{x} (q)$ are $L^{1}$ on $q$-space for any $x\in \mathcal{M}$, and so by eqn. (\ref{H})
$|(f\star g)(x)| <\infty,$ provided $\alpha(q_1,q_2)$ is bounded.
As another example, if $f(x)$ and $g(x)$ are $L^1$ on real space, then by the Riemann-Lebesgue Lemma $|\tilde{f}_{x}(q)|$ and $|\tilde{g}_{x}(q)|$
are bounded, and $|(f\star g)(x)|$ will be bounded when $\int \mathrm{d}q_{1}\,\mathrm{d}q_{2}\, \exp \,[\alpha(q_1 ,q_2 ) ]$ is bounded.
However this last condition is not readily achievable. This is due to ``unitality ridges'' in $\alpha(q_1, q_2)\,$: those places where at least one of $q_1, q_2$ vanishes and the other one is arbitrary, 
fixing $\exp \alpha =1$ there,
because of the unitality condition on the $\star$-product. These simple convergence criteria illustrate that the IKT representation for $(f\star g)(x)$
is considerably more mathematically robust than its DDT ancestor, which must converge as a power series. \\

The IKT still lacks a specific choice for the 2-cocycle $\beta(q)$.  For instance $\beta_{\nu}(q) = - k\, \| q \|^{\nu}$ with positive real constant $k$ for $\nu=2,4$ leads to 
$\alpha_{\nu}(q_1 ,q_2 )$ having indefinite sign. To keep $\exp\,( \alpha) \le 1$ means $\beta(q_1 + q_2 ) \ge \beta(q_1 ) + \beta(q_2 )$, and then one is motivated by the triangle 
inequality to try $\nu=1.$ But for $0<\nu\le 1,$ $\beta_{\nu} (q)$ suffers from another defect. For the IKT to be equivalent to a DDT requires $\alpha$ and $\beta$ to be analytic
in $q$-space. This equivalence is important because once a DDT is in hand so is the entire twist Hopf algebra structure. Since Hopf methods were critical in demonstrating that 
$\star$-diffs are inequivalent to standard diffs \cite{DGRv1.0} as well as for producing the desirable (Land of Oz) properties of the Nice Basis, analyticity of $\beta(q)$ is strongly preferable. That rules out $\alpha_{\nu}$ for
 $0<\nu\le 1$. \\
 
 Now an analytic and bounded $\alpha(q_1,q_2 )$ will be explicitly constructed. If $\xi$ and $\eta$ are $N_X$-tuples, the triangle inequality of the standard Euclidean norm $\| \xi\|$ states $0 \le
 \|\xi +\eta\| \le \|\xi \| + \|\eta \|$. Since the function $\tanh (x)$ is analytic and monotonic for all real $x$, one has 
 \begin{align}
 \tanh\, \| \xi +\eta \| & \le \tanh\, \big( \|\xi\| + \|\eta\| \big) = \big[1 + \tanh\,\|\xi\| \cdot \tanh\,\|\eta\|\big]^{-1} \, \big[\tanh\,\|\xi\| +\tanh\,\|\eta\|\big] \nonumber \\
 & \le \tanh\,\|\xi\| + \tanh\,\|\eta\|,
 \end{align}
 where $\tanh\,\|\eta\| , \tanh\,\|\eta\| \ge 0$ were used in the last inequality. Taking the square yields
 \begin{align}
 \tanh^{2}\, \|\xi +\eta\| & \le \tanh^{2}\, \|\xi\| + 2 \tanh\,\|\xi\| \cdot\tanh\,\|\eta\| + \tanh^{2}\, \|\eta\| \nonumber \\
 & \le \tanh^{2}\, \|\xi\| +\tanh^{2}\,\|\eta\| +2,
 \end{align}
 because $\tanh^{2} x \le 1.$ From this inequality the choice
 \begin{equation}
 \beta_{\mathrm{th}}(q) \doteq - \tanh^{2} (k\,\|q\|)\label{beta_th}\\
\end{equation}
for $k$ a real constant, can be seen to obey
\begin{align}
\beta_{\mathrm{th}} (q_1 + q_2 ) & \ge \beta_{\mathrm{th}} (q_1) +\beta_{\mathrm{th}}(q_2) -2 \quad\text{and} \\
\alpha_{\mathrm{th}}(q_1 ,q_2 ) & = \beta_{\mathrm{th}} (q_1)+\beta_{\mathrm{th}} (q_2) - \beta_{\mathrm{th}} (q_1 + q_2) \le 2. \label{alpha_th_Bnd}
\end{align}
This choice of $\beta$ also implies: (a) Unitality $\beta_{\mathrm{th}}(0)=0,$ (b) Reality $\beta_{\mathrm{th}}(q) = \beta_{\mathrm{th}}(-q) \in \mathbb {R}$, and (c) Commutativity
 and Associativity assured by the first equality in 
(\ref{alpha_th_Bnd}). $\alpha_{\mathrm{th}}$ is bounded analytic everywhere in $q$-space, and its mathematical simplicity is also appealing. \\

The constant $k$ controls the nonlocality scale $\xi_c$ of the theory. In fact physics for momenta approaching or exceeding $\xi_c ^{-1}$ (or $\| q\| \gtrsim k^{-1}$) is essentially unknown.
But it is noteworthy that if $\beta(q)$ were to approach zero as $\|q\| \gg k^{-1}$, then one would revert to the dot product at large momenta, which would be unphysical for this kind of nonlocal theory.
$\beta_{\mathrm{th}}$ by contrast saturates in that limit, capturing that intuition. Nevertheless there remains a great deal of freedom of choice or ambiguity for $\beta$ and its large momentum behavior in
particular, which corresponds to proper lengths much smaller than $\xi_c$. But at those short scales, perhaps even smaller than the Planck length, classical DGR is no longer expected to be an accurate
depiction of Nature since strong quantum spacetime fluctuations play a crucial role. So as far as classical phenomena at lower momenta are concerned, the choice of $\beta$ 
might as well be made on the basis of mathematical convenience and suitability until there is input from measurements. \\

It is vital for subsequent calculations that  the IKT actually possesses a Nice Basis, otherwise one cannot even construct a well-behaved $\star$-covariant derivative. This requires that the IKT's $\star$-product be an Abelian DDT,  which is assured to have a 
Nice Basis. \cite{Aschieri_2}\cite{Schenkel}
We now illustrate how to go from an IKT to a DDT using $\alpha_{\mathrm{th}}$, first starting at low momenta $\| p \| \ll \xi_{c}^{-1}$ and taking $N_{X}=1$ for simplicity.
Since the single generator $X$ arises from GM matter fields, which typically posses momenta $\| p \| \ll \xi_{c}^{-1}$, we may take $X^{\lambda} [\sigma(x,t)]\simeq X^{\lambda} (x),$ independent
of $t,$ so that $\sigma^{\mu}(x,t)\simeq x^{\mu} + t\,X^{\mu}(x)$. Call this the non-recursive gradient approximation; it ignores variations of $X(x)$ over the non-locality scale $\xi_c.$ 
Expanding both $f,g$ in Taylor series about $x$ one has
\begin{align}
(f\star g)(x) & \simeq (2\pi)^{-2} \int \mathrm{d}q_{1}\,\mathrm{d}q_{2}\,\mathrm{d}t_{1}\,\mathrm{d}t_{2}\,\exp\big[ -iq_{1}t_{1}-iq_{2}t_{2}+\alpha(q_{1},q_{2})\big]  f[x^{\lambda} + t_{1}\,X^{\lambda}(x)]
\nonumber \\
& \quad\quad \times g[x^{\lambda} + t_{2}\,X^{\lambda}(x)]  \nonumber\\
& = (2\pi)^{-2} \int \mathrm{d}q_{1}\,\mathrm{d}q_{2}\,\mathrm{d}t_{1}\,\mathrm{d}t_{2}\,\exp\big[\alpha(q_{1},q_{2})\big]\,\sum_{n,m=0}^{\infty} (n!\,m!)^{-1}\nonumber \\
& \quad\quad \times\Big\{ (t_{1}^{n} t_{2}^{m}) \exp\big[ -iq_{1}t_{1}-iq_{2}t_{2}\big] \Big\} 
\Big[ (X^{\lambda}\cdot\partial_{\lambda})^{n}\, f |_{x} \Big] \, \Big[ (X^{\kappa}\cdot\partial_{\kappa})^{m}\, g |_{x} \Big]. \nonumber
\end{align}
Noting $\{\cdots \} =(i\,\partial /  \partial q_{1})^{n} \, (i\,\partial / \partial q_{2})^{m} \,\exp (-iq_{1}t_{1}-iq_{2}t_{2})$, and then integrating by parts leads to
\begin{align}
&(f\star g)(x)  \simeq  \nonumber \\
& \int \mathrm{d}q_{1}\,\mathrm{d}q_{2}\, \exp\big[\alpha(q_{1},q_{2})\big]\, \sum_{n,m=0}^{\infty} \frac {i^{n+m}} {n!\,m!} \, \Big[ \Big(\frac{\partial}{\partial q_{1}}\Big)^{n} \delta(q_1) \,
\Big(\frac{\partial}{\partial q_{2}}\Big)^{m} \delta(q_2) \Big] \Big[ (\mathcal{L}_{X})^{n} f \Big|_x\,\Big] \Big[ (\mathcal{L}_{X})^{m} g \Big|_x\,\Big] \nonumber \\
& = \sum_{n,m=0}^{\infty} \frac {(-i)^{n+m}} {n!\,m!} \, \Bigg\{ \Big( \frac {\partial}{\partial q_{1}} \Big)^{n} \, \Big( \frac {\partial}{\partial q_{2}} \Big)^{m}\,\exp[\alpha(q_1 ,q_2 )] \Bigg\}_{q_1 = 0 =q_2} 
 \Big[ (\mathcal{L}_{X})^{n} f \Big|_x\,\Big] \Big[ (\mathcal{L}_{X})^{m} g \Big|_x\,\Big]. \label{K1}
 \end{align}
Also
 \begin{align}
& \alpha_{\mathrm{th}} (q_1 ,q_2 )   = - \tanh ^{2}(kq_1) - \tanh ^{2}(kq_2) + \tanh^{2} (k(q_1 + q_2 )) \nonumber \\
 &\quad\quad\quad\quad\;\,\simeq \,k^2 \, [(q_1 +q_2 )^2 -(q_1 )^2 - (q_2 )^2] = (2k^2)\, q_1 q_2, \quad\text{and} \nonumber \\
 & \exp \, [\alpha_{\mathrm{th}}(q_1,q_2)]  \simeq \sum _{p=0}^\infty \frac {1}{p!} (2k^2\, q_1 q_2)^{p} \quad \text{for } |kq_1 |,|kq_2 | \ll 1. \label{Lhotse}
 \end{align}
Every term of the last expression in (\ref{Lhotse}) has the same number of $q_1$ factors as $q_2$ factors, so $n=m=p$ in eqn. (\ref{K1}).
 Noting
 \begin{equation}
 (\partial / \partial q_1)^{p} (\partial / \partial q_2)^{p} \exp [\alpha_{\mathrm{th}} (q_1,q_2)]_{(q_1=0=q_2)} = \sum _{p=0}^{\infty} \frac {(p!)^2}{p!}\,(k^2)^{p} = 
\sum _{p=0}^{\infty} (p!)\, (k^2)^{p}, \nonumber
\end{equation}
one finally obtains
\begin{align}
(f\star g)(x) & \simeq \sum_{p=0}^{\infty} \frac{(-i)^{2p}} {p!} (2k^2)^{p} [(\mathcal{L}_{X} )^{p} f |_{x}\,]\,[(\mathcal{L}_{X} )^{p} g |_{x}\,] \nonumber \\
& =\exp\big[ -2k^2 \mathcal{L}_{X} \otimes \mathcal{L}_{X}\big] \triangleright (f\otimes g) (x), \label{DDT_Eq}
\end{align}
which is the familiar DDT eqn. (\ref{DDT}) for a single generator $X$ with DDT expansion parameter $\lambda = 4 k^2.$ 
This approximation is called non-recursive since the final result (\ref{DDT_Eq}) uses the standard Lie action: 
Aside from any possible $\star$-products that might appear within $X,$ no other $\star$-products occur. 
As a bonus, the IKT offers some qualitative insight into the $\star$-product:
It portrays a clear picture of the how $(f\star g)(x)$ gathers information about $f$ and $g$ using the vector generator congruences
 $X_A$ via the displacement functions $\sigma$ on the TSM containing $x$.
Such an explicit geometric interpretation is lacking in the differential DDT version of the twist eqn (\ref{DDT}). 
It might be useful to the reader to keep this description in mind when
specific physical situations are discussed later. The relationship between the DDT and IKT form of this modified theory of gravity is analogous to the differential and integral forms of Maxwell's equations,
with the TSM's  generator congruences roughly playing the role of lines of force.
\\

To what physical regime does the non-recursive gradient approximation correspond? The small parameter $\varepsilon$ of this expansion is
$\varepsilon\sim\xi_C^{2}/L_C \Lambda_C$ in the rest frame of GM matter, where $L_C$ is the spacetime curvature radius and $\Lambda_C$ is the GM 
particle's Compton wavelength. Numerically,
\begin{equation}
\varepsilon \simeq (10^{-56}) (L_C/10^{11}\,\mathrm{m})^{-1} \,(m/m_N)\, (\xi_c/10^{5} L_P)^{2},
\end{equation}
where $m_N \sim 1\,\mathrm{GeV}$ is the nucleon mass and $L_C\sim 10^{11}\,\mathrm{m}$ at the Earth's surface. Near a black hole's event horizon with $L_C \sim 10\,\mathrm{km}$,
$\varepsilon\sim10^{-49}$ for $m=m_N.$ So the non-recursive gradient approximation is astrophysically well grounded for particle momenta much less than $\xi_c^{-1}.$\\

What about a more general case where one does not take the non-recursive gradient approximation and allows possibly up to two generators? By $\star$-product unitality
$\beta(0)=0\,,$ and by reality $\beta(q)=\beta(-q)\in\mathbb{R}$, hence for analytic $\beta$ one can write
\begin{equation}
\beta(q) =\sum_{p=1}^{\infty} A_{p} \| k q\| ^{2p},
\end{equation}
where $A_P\in\mathbb{R}, \,k>0, \,q\in\mathbb{R}^{N_X}.$
Then
\begin{align}
(f_1\star f_2)(x) & =(2\pi)^{-2N_X} \int \mathrm{d}q_{1}\,\mathrm{d}q_{2}\,\mathrm{d}t_{1}\,\mathrm{d}t_{2}\,\exp\big[ -iq_{1}t_{1}-iq_{2}t_{2}\big] \exp \Bigg\{ -\sum_{p=1}^{\infty} A_{p} \,k^{2p} \nonumber\\
& \times\Big[ [(q_1)^{2}]^{p} + [(q_1)^{2}]^{p} - [(q_1 + q_2 )^2]^{p} \Big] \Bigg\} \,f_1 [\sigma(x,t_1)] \,f_2 [\sigma(x,t_2)],
\end{align}
with $(q_j)^2 \doteq \sum_{A} q_{j}^{A} \,q_{j}^{A};$ $j=1,2$ correspond to $f_1 ,f_2\,  ;$ and $A$ is a generator label.
Expand the $\exp$ into a sum of products of powers of the $q_{j}^{A}$, and then translate the powers of $q_{j}^{A}$ into powers of $i(\partial /\partial t_{j}^{A})$ acting on 
$\exp (-i q_j t_j)$, next integrate by parts to transfer those powers of $(\partial/\partial t_{j}^{A})$ onto $f_j[\sigma(t_j)]$ for $j=1,2$ respectively. 
Then take for each fixed $j=1,2$
\begin{align}
\big( \partial /\partial t_{j}^{A} \big) f_j [\sigma(x,t_j)] & \simeq X_{A}^{\kappa}\big(\sigma(x,t_j)\big) \cdot (\partial_{\kappa} f_j)|_{\sigma(x,t_j)}\label{K3} \\
& =[\mathcal{L}_{X_A} \, f_j ] \big(\sigma(x,t_j)\big) , \label{K2}
\end{align}
the standard Lie action of $X_A$ on $f_j$  evaluated at $\sigma(x,t_j).$ Eqn. (\ref{K2}) is a non-recursive (NR) approximation since the $\star$-chain rule would require $\star$ rather
 than $\cdot$ to appear in the middle 
of the RHS of (\ref{K3}),  with recursion one would instead obtain $[X_{A}^{\kappa} \star (\partial_{\kappa} f_j ](\sigma).$ This converts factors of $q$'s into powers of standard Lie derivatives.
After all the dust settles one is left with
\begin{equation}
(f\star g)(x) |_{\mathrm{NR}} = \sum _{p=0}^{\infty} \sum_{l=0}^{p} \sum_{A,B =1}^{N_X} A_{p,l}\, k^{2p} \,(X_A \otimes X_A ) \big[ (X_B)^2 \otimes \hat{1} + \hat{1} \otimes (X_B)^2 \big] ^{p-l}
\triangleright (f\otimes g) |_{x}\; ,
\end{equation}
where $\hat{1}$ is the identity operator $\mathcal{L}_{\hat{1}} f |_x \doteq  f(x)$, $X \triangleright f$ is shorthand for the standard Lie action $\mathcal{L}_{X} f,$
and $A_{p,l}$ are real numerical constants. One can view the $\hat{1}$ operator as a ``do nothing'' generator. It Lie-commutes with any smooth vector field $Z,$ because $[\hat{1}, Z] \triangleright f =
(\mathcal{L}_{\hat{1}}\mathcal{L}_{Z}-\mathcal{L}_{Z}\mathcal{L}_{\hat{1}}) f =(\mathcal{L}_Z - \mathcal{L}_Z ) f =0.$ $\hat{1}$ generates the ``go nowhere'' (null) congruence: 
\begin{align}
\mathrm{d}\sigma_{\hat{1}}^{\mu} / \mathrm{d} t & =0 \\
\sigma_{\hat{1}}^{\mu} (x,t) & =x^{\mu}.
\end{align}
It also trivially idempotent, $\hat{1}^2=\hat{1}.$ So if we enlarge the collection of generators $\{X_A\ : A= 1, \ldots , N_X \}$ by including $\hat{1}$, and call the augmented result 
$\{Y_A\ : A= 1, \ldots , N_X +1\}$ the extended generator set, we can write
\begin{align}
(f\star g)(x) |_{\mathrm{NR}} = \sum _{p=0}^{\infty} \; \sum_{p_1 ,p_2 ,p_3 =0}^{p}  & \tilde{A}(p_1 ,p_2 ,p_3 )\, k^{2p} \,
\Big[ \big(\sum_{A} Y_A^2 \big)^{p_1} \otimes \big(\sum_{B} Y_B^2 \big)^{p_2}\Big] \Big[ \sum_{C} Y_C \otimes Y_C \Big]^{p_3} \triangleright \nonumber \label{L1}\\
& (f \otimes g) |_x \doteq \mathcal{F} \triangleright (f\otimes g) |_x \; .
\end{align}
Here $Y_A ^2 = (Y_A) (Y_A)$ as a Lie operator, and $\tilde{A}(p_1, p_2 ,p_3 ) = \tilde{A}(p_2, p_1 ,p_3 )$ are real-valued. In the non-recursive (NR) approximation the $Y_A$ act on functions  such as $f$ via the standard Lie action,
and in a recursive twist, the $Y_A$ act as $Y_A \triangleright_{\star} f |_x \doteq \mathcal{L}_{Y_A}^{\star} f |_x \doteq [Y_A ^{\mu} \star (\partial_{\mu} f)](x).$
Eqn. (\ref{L1}) is a formal power series for the twist with expansion parameter $k^2.$ Within the NR approximation, this is a Drin'feld twist by definition as a formal power series \cite{Schenkel} using the 
\emph{standard} Lie derivative since it is invertible:
$\mathcal{F}= \hat{1}\otimes \hat{1} + \mathcal{O}(k^2)$, associative and unital. One also can see that a recursive twist is no longer formally a DDT, and is defined only as an IKT,
 because it is equivalent to utilizing a non-standard Lie derivative that recursively employs the $\star$-product. Such a twist generally loses the Hopf algebra structure that the DDT possesses. \\

To illustrate the utility of the IKT, one can introduce $\star$-distributions, such as the ``$\star\,\delta$-function,'' $\delta_{z}^{\star}.$ The action of the $\star\delta$-distribution on some
$f\in C^{\infty}(\mathcal{M})$ is defined by replacing the standard product by $\star$ in the usual definition; i.e., 
\begin{align} 
\delta_{z}^{\star}\, \triangleright f \doteq & \int_{\mathcal{M}} \mathrm{d}^{d} x \, (\delta_{z} \star f)(x)  = (2\pi)^{-2N_X} \int_{\mathcal{M}} \mathrm{d}^{d} x \int
\mathrm{d}^{N_X} q_{1}\, \mathrm{d}^{N_X} q_{2} \,\mathrm{d}^{N_X} t_{1} \, \mathrm{d}^{N_X} t_{2} \,\times \nonumber \\
& \times\exp \big[-i q_{1} t_{1} - iq_{2} {t}_{2} + \alpha(q_1 ,q_2 ) \big]\,
\delta_{z}\big[\sigma(x, t_1 )\big] \,f\big[\sigma(x,t_2 )\big], \label{Delta_Int}
\end{align}
where $\delta_{z} [\cdot]$ is the standard $d$-dimensional $\delta$-function (distribution) based at $z\in \mathcal{M},$ and $d= \dim (\mathcal{M}).$ Choosing $\alpha = \alpha_{\mathrm{th}}$,
and again taking the non-recursive gradient approximation with $\tanh x\simeq x$ discussed earlier, reduces (\ref{Delta_Int})  to a Gaussian integral, 
\begin{equation}
\delta_{z}^{\star}\, \triangleright f \simeq \Big(\frac {\pi}{2k^2 }\Big)^{N_{X}/2} \int _{\mathrm{TSM}(z)} \mathrm{d}^{N_X} \tau\,\exp [-\tau ^2 /8k^2 ] \, f\big[\sigma(z,\tau) \big], \label{Delta_Gaussian}
\end{equation}
a Gaussian-weighted average of $f$ over the $N_X$-dimensional TSM$(z),$ centered at $z.$ Physically this means one can only probe fields ($f$  in eqn. (\ref{Delta_Gaussian})) with Gaussian
 ``fuzziness'' over a proper distance $\sim\xi_c$ 
from $z$ on TSM$(z)$, but with arbitrary (pointlike) precision in the orthogonal directions. This has important physical consequences, to be discussed in later sections. 
Still using $\alpha = \alpha_{\mathrm{th}}$ but without approximating $\tanh x\simeq x$, one has for $N_X =1$
\begin{equation}
\delta_{z}^{\star}\, \triangleright f \,|_{N_X =1}  = k^{-1} \int_{-\infty}^{+\infty} \mathrm{d} \tau \,\mathrm{d} Q\,\exp\Big[ -i \Big( \frac {Q}{k}\Big) \tau - 2 \tanh ^{2} Q \Big] \, f\big[ \sigma(z, \tau) \big]\, ,
\end{equation}
which is difficult to express in terms of standard functions. \\

\section{Simple Dynamics and $\star$-field equations}

Starting from a $\star$-action $S$ consisting of field components starred together to form a $\star$-scalar, one wishes to derive classical $\star$-field equations by varying the fundamental degrees of freedom.
These degrees of freedom will include the GM scalar (fermionic) field $\phi$ ($\psi$), the metrical variables $g_{\mu\nu}$ (tetrad $e^{\mu}_{I}$), and all the standard model non-gravitational fields. 
Because the expressions for  the $\star$-product (\ref{New_STAR}), (\ref{Congruence_ODE}), and (\ref{Congruence_IC}) are dependent on the twist generators $X_A,$ and those
 are derived from $\phi$ or $\psi$ by model equations such as (\ref{XDPhiN}) or (\ref{XFCN}), respectively, $*(X_A)$ is dynamical. This differs from the undeformed case where the dot product is non-dynamical.
 The model equations for the composite object $X$  include $\star$ to assure it transforms as a $\star$-vector under $\star$-coord reparams. This makes the model equations for $X$ 
 take the self-consistent (sc) form
 \begin{equation}
 X^{\mu} = X_{\mathrm{sc}}^{\mu} [ \Psi , *(X) ], \label{XSC} 
 \end{equation}
 with $\Psi$ being all the fundamental fields comprising $X.$
 That is, given any set of fundamental fields $\Psi$, such as $\psi$ together with the tetrad,
which may not necessarily satisfy the field equations, one seeks a fixed ``point'' (really a field configuration) of the iterative map
\begin{equation}
X^{(n+1)} = X_{\mathrm{sc}} \Big[ \Psi, \star\big(X^{(n)}\big) \Big]. \label{X_Iteration}
\end{equation}
An early version of this ``deformation flow'' was discussed in previous work \cite{DGRv1.0}
and it will be examined in more detail in section 7. Here we will assume that given any classical configuration $\Psi$ 
there is a unique solution $\hat{X} [\Psi]$ of eqn. (\ref{XSC}) which will be called the self-consistent generator associated with configuration $\Psi.$ As a reminder, if there are two generators,
then the twin $X_T$ is calculated from the Abelian condition  $[X_T , \hat{X} ]=0.$ \\

Now consider varying fundamental fields about some configuration $\Psi_0$, not necessarily an extremum of the action; and in particular examine $\delta(\star).$
 If one varies a field not contained in $X$, then $\delta(\star) =0,$ and the calculation is straightforward. In the opposite case, generally $\delta X \ne 0$ and $\delta(\star)\ne 0.$
 Suppose we start the variation from some configuration $\Psi_0$ with $X_0 =X_{\mathrm{sc}} [\Psi_0 ]$ and take $\Psi_0 \to \Psi_0 + \delta\Psi$ producing $X_0 \to X_0 + \delta X.$
 Examining the DDT expression eqn. (\ref{DDT}), for the \emph{first order} variation produced by $\delta(\star)\ne 0$ in the action to be overall associative one must have $[X_0 ,\delta X]=0$, where
 $\delta X$ is the first order change in $X.$
 This is because $X_0$ appears together with one instance of $X_0 +\delta X$ within $\delta(\star) = \delta\, [ \, \sum_{n=0} (-\lambda/2)^{n} \, (n!)^{-1} \,  (X^n \otimes X^n ) \, ] ,$ so both $X_0$
 and $X_0 + \delta X$ act as generators and then have to Lie-commute to maintain associativity.  The \emph{ first order} variations $\delta(\star)$ will be constrained by this condition. 
 By contrast, the fully varied twist ($\star$ to all orders in $\delta\Psi$) is always associative, since the DDT 
 \begin{equation}
 \exp\big[ - \frac {\lambda}{2} \theta^{AB} \, \mathcal{L}_{Y_A} \otimes \mathcal{L}_{Y_B}\big] \triangleright (f \otimes g)(x)
 \end{equation}
 is associative, where one $Y_1=X_{\mathrm{sc}}(\Psi +\delta \Psi),$ and if $N_X =2$ then the other $Y_2$ is $Y_1$'s Abelian twin.
  Taking $\delta X =0$ by freezing $X$
 to a self-consistent  $X_0$ at the  configuration $\Psi _0$  is a sufficient solution of $[X_0 , \delta X]=0$ and $\delta(\star)=0$ called ``simple'' dynamics. 
 The constraint 
 $[X_0 ,\delta X]=0$ only affects how one treats $\delta(\star)$ and its internal generator variation $\delta X$ \emph{within the $\star$ operator itself} when varying the action, 
 and this constraint does \emph{not} otherwise restrict the variation of any fundamental fields outside $\star.$
 So far $\Psi_0$ is any given classical field configuration.
 One then seeks a configuration $\Psi_0$ that 
 extremizes the action only under these simple variations, leading to  $\star$-Euler-Lagrange (``simple" equations of motion) for $\Psi_0$.
 The simple variations possess the additional virtue that if one has a Nice Basis $\{e_A\}$ for $\Psi_0$, then that basis will still be Nice for $\Psi_0 +\delta\Psi;$ i.e., simplicity preserves the Nice gauge choice.
 Simple dynamics also means that the field equations for the metrical degrees of freedom and GM fields entering $X$ will not acquire anomalous terms due to $\delta(*) \ne 0$ acting on the $\star$s in 
 the standard matter parts of total Lagrangian. Simple dynamics is studied in this section, and the non-simple case postponed until section 7.  \\

We now focus attention on the torsion-free $\star$-Einstein-Hilbert action in metric variables $g_{\mu\nu}$ whose Lagrangian density is proportional to $|g|^{\star 1/2} \star R.$ Here 
$|g| \doteq  | \det_{\star} g_{\mu\nu} | $ and $(f^{\star 1/2})\star (f^{\star 1/2}) (x) \doteq f(x)$; general rational $\star$-powers are defined analagously.  
Using the same conventions as Weinberg \cite{WGC_ p133_eq_6.1.5}, the definitions of the 
$\star$-Riemann curvature and $\star$-Ricci tensors in a Nice Basis read 
\begin{align}
R^{\lambda}_{\mu\nu\kappa} & \doteq e_{\kappa}(\Gamma^{\star\lambda}_{\mu\nu}) - e_{\nu}(\Gamma^{\star\lambda}_{\mu\kappa} )  + \Gamma^{\star\eta}_{\mu\nu} \star \Gamma^{\star\lambda}_
{\kappa\eta} - \Gamma^{\star\eta}_{\mu\kappa} \star \Gamma^{\star\lambda}_{\nu\eta}\quad\text{and} \\
R_{\mu\kappa} & \doteq g^{\lambda\nu} \star R_{\lambda\mu\nu\kappa}\;, 
\end{align}
and similarly for the $\star$-curvature scalar $R.$
We now sketch how this leads to the $\star$-Einstein field equation using a Nice Basis in combination with simple dynamics.\\

First a few $\star$-calculational tools need to be introduced. Start with
\begin{equation}
\exp_{\star} (f) |_{x} \doteq \sum_{n=0}^{\infty} \frac{1}{n!}\, (f)^{\star n} |_{x}
\end{equation}
and its inverse $\ln _{\star} F = f$ iff $\exp _{\star} f =F.$ It is easy to verify that
\begin{align}
\exp_{\star}( f\star g) & = \exp_{\star} (f) + \exp_{\star} (g)\;\text{and} \\
\ln _{\star}(F\star G) & = \ln_{\star}(F) + \ln_{\star} (G) \; ,
\end{align}
so 
\begin{equation}
\ln_{\star}(F^{*-1}) = (-) \ln_{\star}(F).
\end{equation}
If $M(x)$ is a smooth $4\times 4$ matrix function of position, then when $M\to M+\delta M$
\begin{align}
\delta[ \ln_{\star} \mathrm{det} _{\star} \,M(x) ]& = \ln_{\star} \mathrm{det} _{\star}\,(M+\delta M) - \ln_{\star} \mathrm{det} _{\star} \,(M) \\
& =\ln_{\star} \mathrm{det} _{\star} \, \big[ (M+\delta M)\star M^{\star -1} \big] = \ln_{\star} \mathrm{det} _{\star} \, \big[ 1_{4} +\delta M\star M^{\star -1} \big].
\end{align}
As $\delta M \to 0$ the RHS approaches $\ln _{\star} [ 1 + \mathrm{Tr} M^{\star-1} \star\delta M] \to \mathrm{Tr} [ M^{\star-1} \star \delta M].$ 
Set $\delta M = (\partial_{\lambda} M)\cdot \delta x^{\lambda}$ where $\delta x^{\lambda}\to 0$ is independent of position, so $\delta M = (\partial_{\lambda} M)\star \delta x ^{\lambda}$ as well.
Then 
\begin{align}
(\delta x^{\lambda}) \stackrel{\cdot}{\star} \partial_{\lambda} \big[ \ln_{\star} \mathrm{det} _{\star}\, M(x)\big] & = \mathrm{Tr}  \big[ M^{\star -1} \star (\delta x^{\lambda}  \star \partial_{\lambda} M) \big] \\
& = \mathrm{Tr} \big[ M^{\star -1} \star (\partial_{\lambda} M) \big] \stackrel{\cdot}{\star} (\delta x^{\lambda})\; .
\end{align}
Here $\stackrel{\cdot}{\star}$ indicates either choice of product is valid.
Thus we have the useful result
\begin{equation}
\partial_{\lambda}\big[ \ln_{\star} \mathrm{det} _{\star} \,M(x) \big] = \mathrm{Tr} \big[M^{\star -1} \star \partial_{\lambda} M(x) \big]\; .\label{bbS}
\end{equation}
Applying this with matrix $M$ set to the $\star$-metric tensor, one obtains
\begin{equation}
\Gamma^{\star \mu}_{\mu\lambda} = \frac{1}{2} g^{\mu\rho}\star (\partial_{\lambda} g_{\rho\mu}) = \frac{1}{2} \partial_{\lambda}(\ln_{\star} |g| ) = |g|^{\star -1/2} \star \partial_{\lambda}(|g|^{\star 1/2}),
\end{equation}
where the last equality has used
\begin{equation}
\frac {\mathrm{d} \ln_{\star} (u)}{\mathrm{d}u} = u^{\star -1} \quad\text{for}\; u>0,
\end{equation}
and the $\star$-Chain Rule with $\ln_{\star}\,(f^{\star\alpha}) = \alpha \ln_{\star} f$ for real $\alpha.$
Then for $\star$-vector field $V^{\mu}$
\begin{equation}
V^{\mu}_{\stackrel{\star}{,} \mu} = e_{\mu}(V^{\mu}) + \Gamma^{\star\mu}_{\mu\lambda} \star V^{\lambda} = e_{\mu} (V^{\mu}) + V^{\lambda}\star|g|^{\star-1/2} \star e_{\lambda}( |g|^{\star 1/2}).
\end{equation}
In a Nice Basis $e_{\hat{\mu}}$ where the Lie-derivative is $\star$-Leibniz one finally gets
\begin{equation}
|\hat g|^{\star1/2} \star V^{\hat\mu}_{\stackrel{\star}{,}\hat\mu} = e_{\hat\lambda} \big[ (|\hat g|^{\star 1/2})\star V^{\hat\lambda} \big]\; .\quad  \text {(In a Nice Basis)} \label{bbT}
\end{equation}
\\

Proceeding further along standard lines, one defines the $\star$-stress energy tensor $T^{\mu\nu}$ by the variation
\begin{equation}
\delta S_{\mathrm{matter}} \doteq \frac{1}{2} \int \mathrm{d}^{4}x \, |g|^{\star1/2} \star T^{\mu\nu}\star \delta g_{\mu\nu} \; .
\end{equation}
For simple dynamics the stationarity of the action under $\delta g_{\mu\nu}$ leads to
\begin{equation}
\int \mathrm{d}^{4}x \, |g|^{\star1/2} \star \Big[  \delta g_{\mu\nu}\star(R^{\mu\nu}-\frac{1}{2} R\star g^{\mu\nu} + 8\pi G T^{\mu\nu}) -g^{\mu\nu} \star
(\delta R_{\mu\nu}) \Big] = 0, \quad \text{with} \label{bbU}
\end{equation}
\begin{align}
\delta R_{\mu\nu}  = & \frac{1}{2}\Big\{ g^{\lambda\rho} \star \big[ (\delta g_{\rho\mu})_{\stackrel{\star}{,}\lambda} + (\delta g_{\rho\lambda})_{\stackrel{\star}{,}\mu} - (\delta g_{\mu\lambda})_{\stackrel{\star}{,}\rho} 
\big] \Big\}_{\stackrel{\star}{,}\nu} \nonumber \\
& - \frac{1}{2}\Big\{ g^{\lambda\rho} \star \big[ (\delta g_{\rho\mu})_{\stackrel{\star}{,}\nu} + (\delta g_{\rho\nu})_{\stackrel{\star}{,}\mu} - (\delta g_{\mu\nu})_{\stackrel{\star}{,}\rho} 
\big] \Big\}_{\stackrel{\star}{,}\lambda}\, .
\end{align}
In a fashionable Nice Basis, where coordinate indices all sport carets and $\star$-contraction commutes with $\star$-covariant differentiation, one finds
\begin{equation}
g^{\hat\mu\hat\nu} \star (\delta R_{\hat\mu\hat\nu}) = V^{\hat\nu}_{\stackrel{\star}{,}\hat\nu} - W^{\hat\kappa}_{\stackrel{\star}{,}\hat\kappa} \quad  \text{(In a Nice Basis)}
\end{equation}
for $\star$-vector fields $V, W.$ Then by eqn. (\ref{bbT}) $|g|^{\star 1/2} \star g^{\hat\mu\hat\nu}\star\delta R_{\hat\mu\hat\nu}$ is a total derivative whose contribution to
(\ref{bbU}) vanishes when $\delta g_{\hat\mu\hat\nu} \to 0$ as $x\to\infty.$ As usual, since $\delta g_{\hat\mu\hat\nu}$ is arbitrary and smooth given the fixed Nice Basis choice, one has
\begin{equation}
R^{\hat\mu\hat\nu} -\frac{1}{2} R\star g^{\hat\mu\hat\nu} + 8\pi G\, T^{\hat\mu\hat\nu} = 0.\quad  \text{(In a Nice Basis)}
\end{equation}
However, since this is the vanishing of a $\star$-tensor, it obtains in all coordinate bases, so one can immediately remove the carets as mere fashion statements to arrive at
 \begin{equation}
R^{\mu\nu} -\frac{1}{2} R\star g^{\mu\nu} + 8\pi G\, T^{\mu\nu} = 0, \quad\text{(In Any Basis)} \label{Star_Einstein}
\end{equation}
the simple $\star$-Einstein field equation (using Weinberg's conventions). There are no extra twist terms $\Delta_{\mu\nu}$ like those anticipated in earlier work \cite{DGRv1.0}.
Eqn. (\ref{Star_Einstein}) was derived within simple dynamics $\delta(\star) =0,$ where the variations $\delta g_{\mu\nu}$ 
about the stationary action configuration remain \emph{unconstrained}
while the twist generator(s) and $\star$ remain \emph{frozen} to their action stationary point values. In this sense simple dynamics is only partially consistent. 
It preserves background independence and Lorentz invariance, and
it still allows the twist to possess some partial dynamics through the generators' self-consistent dependence on the field configurations $\Psi_0$ \emph{at} the action's stationary point.
But it does not allow so much freedom that the twist 
dynamics would disrupt orderly mathematical progress, such as when the twist follows the variations. 
Simple dynamics is then a tame ``teenage'' version of the wilder non-simple dynamics discussed in section 7 below.\\

By choosing to work in a Nice Basis, one may use standard methods \cite{Nakahara_p269_70} to derive the first and second $\star$-Bianchi identities
\begin{align}
R^{\hat\kappa}_{\;\,[\hat\lambda\hat\mu\hat\nu]} & = 0\quad\text{and} \\
R_{\hat\lambda\hat\mu[\hat\nu\hat\kappa\stackrel{\star}{,}\hat\eta]} & =0. \label{bbW}
\end{align}
And again, by the now familiar argument, the carets are superfluous. 
From eqn. (\ref{bbW}) by similar means one can verify that 
\begin{align}
(R^{\mu\nu} -\frac{1}{2} g^{\mu\nu} *R)_{\stackrel{\star}{,}\mu} & =0\quad\text{ and then}\\
T^{\mu\nu}_{\;\;\;\;\stackrel{\star}{,}\nu} & =0,
\end{align}
reassuring us that $\star$-4-momentum conservation is alive and well in simple dynamics. \\

\section{$\star$-gauge theory and friends} 

With a little work matter gauge theories can be exported from undeformed manifolds to commutative $\star$-manifolds. We begin with an intuitive introduction. 
Consider the $\star$-gauge transformation  of matter fields $\psi_l (x)$ 
\begin{equation}
\delta\psi_l (x) = i\epsilon^{\alpha} (x) \star (\hat{t}_{\alpha} )_{l}^{\;m}\, \psi _m (x), \label {G1}
\end{equation}
where $\alpha$ labels some set of \emph{position independent} Hermitian Lie algebra generators $\{ \hat{t}_{\alpha} \},$  ${}_{l}^{\;m}$ denotes the $l$-th row and $m$-th column
matrix element, and $\epsilon^{\alpha}(x)$ is a smooth set of real-valued gauge transformation parameters \cite{W_QFT2_p2}. Normally one would seek to cancel the second term on the RHS of
\begin{equation}
\delta \big( \mathcal{L}_{e_{\mu}} \,\psi_{l} (x) \big) = i \epsilon^{\alpha}(x) \star (\hat{t}_{\alpha})_{l}^{\;m}\, (\mathcal{L}_{e_{\mu}} \psi_{m} )(x) + i(\mathcal{L}_{e_{\mu}} \epsilon^{\alpha})
\star (\hat{t}_{\alpha} )_{l}^{\;m} \psi _m (x) \label{G2}
\end{equation}
by introducing a gauge potential field $A_{\mu}^{\beta}$ having the gauge transformation 
\begin{equation}
\delta A_{\mu}^{\beta} = \mathcal{L} _{e_{\mu}} \epsilon^{\beta} + i \epsilon^{\alpha} \star (\hat{t} ^{A}_{\alpha})^{\;\beta}_{\gamma} A^{\gamma}_{\mu} \;  , \label{G3}
\end{equation}
with generators in the adjoint representation $\hat{t}^{A}_{\alpha}$ being defined by $[\hat{t}_{\alpha} , \hat{t}_{\beta} ] = i C^{\gamma}_{\alpha\beta} \hat{t}_{\gamma}\doteq 
- (\hat{t}^{A}_{\beta})^{\;\gamma}_{\alpha} \,\hat{t}_{\gamma}.$
However in expression (\ref{G2}) the Leibniz rule has been used, and the Lie derivative $\mathcal{L}$ is generally not $\star$-Leibniz, with the exception of 
an Abelian twist in the Nice Basis $e_{\hat{\mu}}.$ Choose such a Nice Basis and drop the associated carets on indices. By defining the $\star$-gauge covariant derivative
 by analogy with the standard version as 
 \begin{equation}
 \Big(D^{\star}_{\mu} \,\psi(x) \Big)_l \doteq (\mathcal{L}_{e_{\mu}} \psi_l ) (x) - i A^{\beta}_{\mu} \star (\hat{t}_{\beta})^{\;m}_{l} \,\psi_{m} (x) \; , \label{G4}
 \end{equation}
 it is straightforward to derive
 \begin{equation}
 \delta\big[ D^{\star}_{\mu} \,\psi \big]_l = i\epsilon^{\alpha} \star (\hat{t}_{\alpha})^{\;n}_{l} \,\big(D^{\star}_{\mu} \psi \big) \label{G5}
 \end{equation}
 as one desires. Since this expression may be written as the vanishing of a $\star$-vector, it is a coordinate basis independent statement.
 It is necessary to have $\delta(\star)=0$; i.e., the gauge transformation $\delta$ does not affect the $\star$ coming from the second term on the RHS of (\ref{G4}). This means the \emph{twist generators}
 $X_A$ must be gauge invariants (singlets), which was demonstrated in \cite{DGRv1.0} as a consequence of overall gauge invariance of the $\star$-action. 
 As was shown in that work, $X$ can then be constructed either from a $U(1)_{GM}$ gauge current (not necessarily the electromagnetic $U(1)$) or from a gauge-less current.
 However, that reasoning does \emph{not} imply that the twist producing GM matter \emph{fields}  are only  $U(1)_{GM}$
 gauge interacting or gauge-less. GM matter may carry other gauges, but non-$U(1)$ gauge \emph{currents } cannot enter the twist generator(s). \\
 
 One can go further and adapt the fiber bundle formalism for gauge theories to $\star$-commutative manifolds. At first this might seem surprising because the matrix gauge (structure) 
 \emph{group} $G$ is no longer the same over every point  $p$ on the base manifold $M.$ Instead of the gauge group binary multiplication operation being position $p$ independent, it becomes 
 position dependent since the twist generators are, and $G$ is deformed to $G_p.$ However, the Lie algebra,  $\mathrm{Lie} \, G,$ remains position independent, and that saves $\star$-gauge theory.  So for example, a group element
 $g(p)\in G_p $ (a smoothly varying matrix-valued function of $p$) may be computed as
 \begin{equation}
 g(p) = \exp _{\star} \, [ v^{\alpha}(p)\, \hat{t}_{\alpha} ]\, ,\label{G6}
 \end{equation}
 where the $\hat{t}_{\alpha}$ are generators of the same position independent Lie algebra $\mathfrak{g}_c = \mathrm{Lie} \,G$ as the undeformed case, and $v^{\alpha} \in C^{\infty}(M).$
 The subscript $c$ is to remind one that the Lie algebra is constant, and $p$ on $G_p$ recalls that group multiplication is position dependent. Clearly $G_c$ is a subgroup of $G_p$ where all the 
 $v^{\alpha}$ are constants. In fact expanding (\ref{G6}) near the identity matrix $e_G$ one has
 \begin{equation}
 g(p) = e_G + v^{\alpha}(p)\,  \hat{t}_{\alpha} +(v^{\alpha} \star v^{\beta})(p) \, (\hat{t}_{\alpha}\, \hat{t}_{\beta} ) + \cdots\; .  \label{G7}
 \end{equation}
 This shows the first order (group tangent space at $e_G$) piece of $g(p)$ is unaffected by $\star\ne\cdot$  and so $\mathrm{Lie} \, G_p = \mathrm{Lie} \, G = \mathfrak{g}_c$.   \\
 
 Formally \cite{Nakahara_Chap_10} one introduces a $\star$-principal bundle $P(M, G_p ,\star(X))$ over the base manifold $M$ with left and right $\star$-group actions, local $\star$-sections $\sigma(p),$ 
 vertical $V_u P$ and horizontal $H_u P$ decompositions of the tangent bundle $T_u P$ for $u \in P.$  This $\star$-fiber bundle is then equipped with $\omega ,$ a $\mathfrak{g}_c$-valued $\star$-one-form 
 on $P$ as a $\star$-Ehresmann connection.
Finally, $\omega$'s  $\star$-pull-back to $M$ is the Lie algebra-valued $\star$-gauge potential one-form on $M,$ $A(p)\in \mathfrak{g}_c\otimes T^{*} M.$ Here the detailed derivations are omitted and
 the salient results summarized. The derivations mostly follow \cite{Nakahara_Chap_10} , with some tricky places where Lie derivatives have to be suitably 
 replaced by exterior derivatives on the $\star$-manifold $M$, along the lines of Section 2 above. \\

Given two local sections $\sigma_1 (p)$ and $\sigma_2 (p)$ with 
\begin{equation}
\sigma_2 (p) = \sigma_1 (p) \star g(p), \label{Star_GXform}
\end{equation} 
a $\star$-gauge transformation, then
\begin{equation}
A_2 (p) = (g^{\star -1} \star A_1 \star g)(p) + ( g^{\star -1} \star \mathrm{d}_{M} \, g) (p)\, ,\label{A_gauge_xform}
\end{equation}
simply the starred version of the standard result.  Similarly one can define unique $\star$-horizontal lifts of smooth curves $\gamma(t): [0,1]\to M$ to $\tilde{\gamma}(t)$ on $P,$
and any other horizontal lift $\tilde{\gamma}'(t)$ of $\gamma(t)$ is related to $\tilde{\gamma}'(t)$ by $ \tilde{\gamma}'(t) = \tilde{\gamma}(t) \cdot g$ for some $g\in G_c $, the constant group.
Consequently one may introduce unique (gauge) $\star$-parallel transport along curves and then $\star$-gauge covariant derivatives $\mathrm{D}.$ In particular one can define the $\star$-gauge curvature
$\star$-2-form on $P,$ $\Omega\in \Omega^2 (P)\otimes \mathfrak{g}_c$ as the $\star$-covariant derivative $\Omega \doteq \mathrm{D}\,\omega.$ This satisfies the $\star$-Cartan structure equation
\begin{equation}
\Omega \doteq \mathrm{D}\,\omega = \mathrm{d}_P \,\omega + \omega\wedge_{\star} \omega\, . \label{Star_CS}
\end{equation}
Finally, the local $\star$-field strength $\mathcal{F}, $ a $\mathfrak{g}_c$-valued $\star$-2-form on $M,$ is derived from $\Omega$ as the $\star$-pullback to $M,$ 
$\mathcal{F}\doteq \sigma^{(\star)*} \, \Omega$ by local sections. $\mathcal{F}$ obeys the $\star$-gauge transformation property
\begin{equation}
\mathcal{F}_2 = g^{\star -1} \star \mathcal{F}_1 \star g ,
\end{equation}
analagous to (\ref{A_gauge_xform}).  $\Omega$ obeys the $\star$-Bianchi identity $\mathrm{D}\, \Omega =0,$ or in local form
\begin{equation}
\mathrm{D}\,\mathcal{F} \doteq \mathrm{d}_M \, \mathcal{F} +A\wedge_{\star} \mathcal{F} - \mathcal{F}\wedge_{\star} A =0\, . \label{Star_Bianchi_2}
\end{equation}
Taking $\mathfrak{g}_c = \mathfrak{so}(3,1),$ the Lorentz Lie algebra, and for vanishing $\star$-torsion $T^I \doteq \mathrm{d}_M \,e^I +\tilde{\omega} ^I_{\;J} \wedge_{\star}
e^J =0,$ for $\tilde\omega^I _J$ the $\star$-spin connection, then the $\star$-Bianchi identity  (\ref{Star_Bianchi_2}) is readily shown to coincide with the coordinate version 
$R^{\lambda}_{\;\;\nu[\rho\sigma\stackrel{\star}{,}\mu]} =0$ derived earlier.\\

Holonomies $g(t)$ along a smooth curve $\gamma(t)$ lying in $M$ with tangent vector $Y$ are constant group $G_c$-valued. They obey the non-starred ODE
\begin{equation}
\frac{\mathrm{d}g(t)}{\mathrm{d}t} = - \langle A,Y\rangle_{\cdot} \cdot g(t)\, , \label{Holonom_ODE}
\end{equation}
with initial condition $g(0) =e_G.$ There are no stars in (\ref{Holonom_ODE}) because the holonomy is a one dimensional object, 
most readily shown using a Lie-flat ``fattening" of the path and then the  IKT reduces the $\star$ to
the standard product.
 Holonomies $h_{\gamma}(u) \in G_c$ taken around closed smooth curves based at $p\in M$, 
 $\gamma :[0,1]\to M,\, \gamma(0) = \gamma(1) = p,$
 with $p$ being a projection $\pi(u),$ for $u\in P,$ can be used to define holonomy groups as: 
 $\Phi(u) \doteq \{ g\in G_c | h_{\gamma}(u) = u\cdot g,\; \gamma\in C_p(M)\}, $ $C_p(M)$ being the set of all closed smooth curves in $M$ based at $p.$
 For path-connected $M$ all the $\Phi(u)$ are isomorphic, as usual. However holonomies taken along non-closed paths $\beta$ are of limited utility
 on $\star$-manifolds. This is because under a $\star$-gauge transformation eqn. (\ref{Star_GXform}), the path holonomy transforms according to
 \begin{equation}
 G_c \owns h_c (\beta) \mapsto g\big(\beta(0)\big) \star h_c \big(\beta\big) \star g\big( \beta(1)\big), \label{GI}
 \end{equation}
 where the subscript $c$ emphasizes that the object takes values in the constant group $G_c .$ But the RHS of (\ref{GI}) is ill-defined for non-closed paths
 $\beta(0) \ne\beta(1),$ and group $\star$-multiplication by $g\big(\beta(0)\big)$ is \emph{not}
 the same as by $g\big(\beta(1)\big)$ for position dependent $g.$ Even though $h_c(\beta)\in G_c,$ it is not generally possible to form 
 the \emph{two} $\star$-products on the RHS of eqn. (\ref{GI}) consistently. Thus only for closed paths are holonomies $\star$-gauge covariant. Consequently calculations employing $\star$-gauge structures
  with open paths, such as $\star$-spin networks \cite{Rovelli_QG}, will not maintain gauge invariance. Loop-based methods like Wilson loops or the loop technology of Gambini and Pullin \cite{Gambini_Pullin}, 
  however, are free from this defect
  when extended to $\star$-manifolds. Unfortunately, the Gambini-Pullin loop technology is of limited use for quantizing DGR because the $\star$-constraints differ too much from classical GR,  as will be discussed in section 7.\\
  
  \section{Banishment of photon and graviton $\star$-ghosts from Minkowski spacetime}
  
  We start by demonstrating that $\star$-propagation of electromagnetic radiation remains ghost-free in Minkowski spacetime.
  The action in generally curved spacetime reads
 \begin{equation}
 S_{\mathrm{EM}}  = -\frac{1}{4}\int_{M} \mathrm{d}^4 x\, |g|^{\star 1/2} \star g^{\mu\lambda} \star g^{\nu\kappa}\star F_{\mu\nu} \star F_{\lambda\kappa}\, ,
 \end{equation}
 with field strength $F_{\mu\nu} = A_{\mu\stackrel{\star}{,} \nu} - A_{\nu\stackrel{\star}{,} \mu}.$ In the absence of torsion, this can be written as
 $F_{\mu\nu} = e_{\nu} (A_{\mu}) - e_{\mu}(A_{\nu}).$ Since $\delta(X)=0=\delta(\star)$ for $\delta A_{\mu} \ne 0$ (no need to assume simple dynamics), in a Nice Basis (carets) and in the absence of current sources one obtains
 \begin{align}
 \delta S_{\mathrm{EM}} =0 & = \int_M \mathrm{d}^4 x\, |g|^{\star 1/2} \star g^{\hat{\mu}\hat{\lambda}} \star g^{\hat{\nu}\hat{\kappa}}\star \big[ e_{\hat{\nu}} (\delta A_{\hat{\mu}}) - e_{\hat{\mu}} 
 (\delta A_{\hat{\nu}} ) \big] \star \big[ e_{\hat{\kappa}} (A_{\hat{\lambda}}) - e_{\hat{\lambda}} (A_{\hat{\kappa}}) \big] \nonumber \\
 & = \int_M \mathrm{d}^4 x\, \delta A_{\hat{\mu}} \star e_{\hat{\nu}} \big[ |g|^{\star1/2} \star g^{\hat{\mu}\hat{\lambda}} \star g^{\hat{\nu}\hat{\kappa}} \star F_{\hat{\lambda}\hat{\kappa}} \big]\; ,
 \end{align}
 where the Nice Basis' $e_{\hat{\mu}}$ $\star$-Leibniz property has be utilized to integrate by parts, and the surface piece where $F$  vanishes has been discarded.
 Hence $e_{\hat{\nu}} \,(\,|g|^{\star1/2} \star F^{\hat{\mu}\hat{\nu}} \,) =0.$ Upon starring this expression with $|g|^{\star-1/2}$ this states $F^{\hat{\mu}\hat{\nu}}_{\quad\stackrel{\star}{,}\hat{\nu}} = 0$ by eqn.
 (\ref{bbT}) in a Nice Basis.  Applying the commutativity of $\star$-contraction and $\star$-covariant derivatives in that basis one can write 
 \begin{align}
& g^{\hat{\mu}\hat{\nu}} \star g^{\hat{\lambda}\hat{\kappa}} \star F_{\hat{\nu}\hat{\kappa}\stackrel{\star}{,}\hat{\lambda}} =0, \quad\text{or} \nonumber \\
& g^{\hat{\lambda}\hat{\kappa}}\star F_{\hat{\mu}\hat{\kappa}\stackrel{\star}{,} \hat{\lambda}} = 0.
\end{align}
Again by by the usual $\star$-tensorality argument, this holds in any basis, so going to the standard Minkowskian $(t,x,y,z)$ coordinate basis  and setting $g^{\lambda\kappa} = \eta^{\lambda\kappa}$
and $\Gamma ^{\star \mu}_{\lambda\kappa} =0$ gives $\eta^{\lambda\kappa} \cdot \partial_{\lambda} \,F_{\mu\kappa} =0$ for flat spacetime and 
\begin{equation}
\eta^{\lambda\kappa} \cdot (\partial_{\kappa}\partial_{\lambda} A_{\mu} - \partial_{\mu}\partial_{\lambda} A_{\kappa} ) =0.
 \end{equation}
 Choosing the gauge $\partial^{\lambda} A_{\lambda} =0$, one arrives at the familiar $[\partial^{\lambda}\partial_{\lambda}] A_{\mu} = 0.$
 So there are no photon ghosts on Minkowski spacetime. This is yet another instance of the well-known fact that deformations in Minkowski spacetime have limited physical consequences. \\
 
 What about on $\star$-curved spacetime? Using the DDT to expand $g^{\lambda\kappa} \star F_{\mu\kappa\stackrel{\star}{,} \lambda} = 0$ it is easy to determine the expansion parameter
 $\varepsilon_{\mathrm{EM}} = \xi_c ^2 /\lambda_{\mathrm{EM}} L_c, $ for electromagnetic wavelength $\lambda_{\mathrm{EM}}$ and spacetime curvature radius $L_c$. 
 The zeroth and first order effect in the $\star$-wave equation is $\mathcal{O}(1+\varepsilon).$
 Taking $\xi_c \simeq 10^5 L_P$,
 $\lambda_{\mathrm{EM}} \sim 1.2 \cdot 10 ^{-15}$ m (a 1 GeV $\gamma$-ray) and $L_c \simeq 10$ km for a astrophysical black hole's event horizon radius leads to $\varepsilon \simeq 8 \cdot 10 ^{-50}.$
 So commutative deformations with curvature are unlikely to play a measurable role in astrophysical electromagnetic wave propagation. $\Gamma$-based effects such as lensing  phenomena
 will be deformed by the same relative amount. \\
 
Gravitational waves about Minkowski spacetime yield a similar null result. This occurs because of the decomposition $g_{\mu\nu} = \eta_{\mu\nu} +h_{\mu\nu}.$ The propagation equation is linear in $h$
and $\star$ drops out. So there are no gravitational ghosts either.\\

Both electromagnetic and gravitational phenomena will undergo significant to severe deformations in the cosmological epoch when $\xi_c \simeq L_c \sim H^{-1}(a)$ at small scale factor $a.$
This is an acausal epoch (ACE)  due to the nonlocal micro-causality violations being non-perturbative then. It is also the epoch when quantum corrections to the twist generators at momentum scale $\xi_c^{-1}$
are non-negligible, and the classical theory becomes inapplicable.\\

\section{ Non-simple dynamics and canonical DGR}

In the previous sections the matter produced twist generator was constructed as a solution of the iterative/recursive process eqn. (\ref{X_Iteration})
\begin{equation}
X^{(n+1)} = X_{\mathrm{sc}} \Big[ \Psi, \star\big(X^{(n)}\big) \Big] ,\label{NSD_DefFlow}
\end{equation}
where one is given some fundamental field configuration $\Psi$, not necessarily a stationary point of the action. For instance, for fermionic GM matter producing
$X$ as a classical $U(1)_{GM}$ gauge current,
\begin{equation}
X^{\mu}_{\mathrm{sc}} \big[ \psi ,e^{\mu}_{I}, \star(X) \big] = \Bigg\langle\binom {\mathrm{Re}}{\mathrm{Im}} \sum_l Q_l \,\bar{\psi}_l \, \star_{X} \, \gamma^I e^{\mu}_{I} \, \star_{X}\, \psi_l \Bigg\rangle
\label{NSD_fermi_eg} \, ,
 \end{equation}
 with  species index $l$ and $U(1)_{GM}$ charge $Q_l.$ Maintaining first order associativity of the action under a variation $\Psi\to\Psi+\delta\Psi$ imposes
 \begin{equation}
 [X, \delta X]=0 \label{NSD_Assoc}
 \end{equation}
 on $\delta(\star).$ To derive the $\star$-Einstein equation (\ref{Star_Einstein}) the simple solution $\delta X =0= \delta(\star)$  of (\ref{NSD_Assoc})  was employed. The more general case is examined here.
 For simplicity, in the rest of this section it is assumed that there is only one twist generator, $N_X =1.$ \\
 
 Does eqn. (\ref{NSD_Assoc}) lead to a consistent set of conditions on the first order $\delta(\star)?$ Imagine three distinct mutually nearby configurations $\Psi_j $ with  $j=1,2,3$ each with an associated 
 self-consistent $X_j.$ Let $\delta X_{ij} \doteq X_j -X_i$ denote the change in $X$ upon varying $\Psi_i \to \Psi_j ,$ so (\ref{NSD_Assoc}) reads $[X_i , \delta X_{ij}]=0.$ What are the allowed paths
 of variations for $\delta(\star)$  in configuration space? Since $\delta X_{12} +\delta X_{23} +\delta X_{31} =0,$ any closed paths are permissible. 
 Similarly if one can travel from $\Psi_1$ to $\Psi_2$, then the reverse journey
 is also permitted.  For three configurations, one has $\delta X_{13} = \delta X_{12} +\delta X_{23},$ and upon applying $[X_i , \delta X_{ij}]=0$ twice one obtains
 \begin{equation}
 [X_1, \delta X_{13}] = [ X_1, \delta X_{12} +\delta X_{23}] = [X_1, \delta X_{23}] =[ (X_1 -X_2 ) +X_2 , \delta X_{23}] = -[\delta X_{12}, \delta X_{23}]. \nonumber
 \end{equation}
 Generally the RHS is non-vanishing. However it is of \emph{second} order in the variations, so the LHS vanishes to first order, which is all we need. That way if the Ordainer of Order allows travel 
 from 1 to 2 as well as from 2 to 3, direct travel from 1 to 3 is also permissible. Thus the condition (\ref{NSD_Assoc}) for $\delta(\star)$ is consistent through first order,
  and there is no path dependence for those variations $\delta(\star).$ \\
  
  From the DDT form of the twist $(f\star_X g)(x) =\exp [ -(\lambda/2) \, X\otimes X] (f\otimes g) (x),$ and using $[X, \delta X] =0$ starting from some initial field configuration $\Psi_0$ 
  to impose first order associativity on some term in the varied action $\delta S =\delta \int\mathrm{d}^4 x\, (f\star_X g)(x),$ one finds
  \begin{align}
  \frac{\delta^{[\star] }} {\delta\Phi(z)} \int \mathrm{d}^4 x\, (f\, {}^{\;\,\star}_{(x)} \, g)(x)
  & = - \frac{\lambda}{2}  \int \mathrm{d}^4 x\, \Big\{ \big[ (\mathcal{L}_{X}^{(x)}\,f) \,\star_{X} \, (\mathcal{L}^{(x)}_{w(x,z)}\, g)\big] (x) 
  + \big[ f\rightleftharpoons g \big] \Big\}, \label{NSD_VarStar} \\
  \text{with}\quad w^{\sigma}(x,z) & \doteq \int \mathrm{d}^4 y\, \frac{\delta^{\star}\,X^{\sigma} (y)} {\delta \Phi(z)} \,{}^{\;\,\star}_{(y)} \, \delta ^{4} _x (y). \label{NSD_w}
  \end{align}
  $w^{\sigma}(x,z)$ are the components of the $\star$-vector field $w(x,z)$ at $x$ for some fixed $z$ coming from the LHS of (\ref{NSD_VarStar}). 
  The notation ${}^{\;\,\star}_{(y)}$ means the $\star$-product is to be evaluated at the point $y;$ this differs from $\star_X$ which specifies the vector field $X$ as the generator of $\star.$
  The variation $\delta ^{[\star]} (f\star g)$ acts \emph{only} on $\star$ but \emph{not} on $f$ or $g.$ It should be distinguished from $\delta^{\star}$ which is simply the $\star$-functional derivative
  defined by:
  \begin{equation}
  \delta X(y) \doteq  \int \mathrm{d}^4 z\,  \frac{\delta^{\star}\,X(y)}{\delta\Phi(z)} \,{}^{\;\,\star}_{(z)} \, \delta\Phi(z).
  \end{equation}
  All the $\star$-products in these expressions are evaluated using $X_{\mathrm{sc}} [\Phi_0 ].$ 
  In particular, $w=0$ for the simple dynamics studied earlier, and then the RHS of (\ref{NSD_VarStar}) vanishes. Eqn. (\ref{NSD_VarStar}) may be more compactly expressed as
  \begin{equation}
 \delta^{[\star]}\,\int  \mathrm{d}^4 x\, (f\star_X g) (x) = -\frac{\lambda}{2} \int  \mathrm{d}^4 x\, \Bigg\{ \Big[ (\mathcal{L}_{X}^{(x)}\, f)  \, {}^{\;\,\star}_{(x)} \, (\mathcal{L}_{\delta X}^{(x)} \,
 g) \Big](x) + \Big[ f\rightleftharpoons g \Big] (x) \Bigg\}. \label{NSD_H}
 \end{equation}
 Which is modest progress because at least it is a $\star$-closed expression for the first order variation $\delta(\star).$
 Notice, however, that that the Lie derivative includes a dot-product: 
 \begin{equation}
 \big( \mathcal{L}_{\delta X}^{(x)} \,g\big) = \delta X^{\sigma}(x) \cdot (\partial g(x)/ \partial x^{\sigma}).
 \end{equation}
 Taking the fermionic example eqn. (\ref{NSD_fermi_eg}) above, for some variation $\delta e^{\mu}_{I}$ of the tetrad one can write
 \begin{equation}
 \delta X^{\sigma} (x) = \big[ \delta e^{\mu}_{I}\,\star\, A^{I\sigma}_{\mu} \big](x).
 \end{equation}
 Now one encounters a mathematical obstacle to deriving the Euler-Lagrange (field) equations
 from $\delta_e S=0$ called non-associative locking. Namely the dot- and $\star$-products are generally mutually non-associative:
 \begin{equation}
 \big\{f\star [g\cdot h]\big\}(x) \ne \big\{ [f\star g] \cdot h\big\}(x). \label{NSD_NAL}
 \end{equation}
 A similar situation was encountered earlier at the end of section 2 in eqn. (\ref{NAL1}). 
 To be compatible with terms where $\delta$ acts directly on $f$ or $g$ rather than on $\star,$ one would like to express the classical stationarity condition $\delta_F S=0$
  under the variation of some field component $F$ in the $\star$-factored form
 \begin{equation}
 \int \mathrm{d}^4 x\, \delta F(x)  \,{}^{\;\,\star}_{(x)} \, \big\{ \cdots \big\}(x) = 0,
 \end{equation}
 and then obtain the field equation $\{\cdots\} (x)=0.$ That is not generally possible because the terms in the integral on the RHS of eqn. (\ref{NSD_H}) are of the form
 \begin{equation}
 h \star \Big[ (  \delta e^{\mu}_{I} \,\star \, A^{I\sigma}_{\mu} )  \cdot (\partial_{\sigma} g ) \Big],
 \end{equation}
 which non-associative locking obstructs from being the sought after form
 \begin{equation}
 (\delta e^{\mu}_{I} \,\star\, A^{I\sigma}_{\mu} ) \star \Big[ h \cdot (\partial_{\sigma} g) \Big]. 
 \end{equation}
 As a result, the change
 \begin{equation}
 \delta^{[\star]}\,\int _M \mathrm{d}^4 x\, (f\star_X g) (x)
 \end{equation}
 with respect to some change in a field $\Psi$  contained in $X$
 \emph{cannot be written} in terms of a $\star$-variational derivative
 \begin{equation}
 \int _M \mathrm{d}^4 x\; \delta\Psi(x)  \, {}^{\;\,\star}_{(x)} \, \{\cdots\}(x)
 \end{equation}
 for non-simple dynamics: That variational $\delta^{[\star]}$ derivative is simply not defined!
 For instance, consider  the $\star$-Poisson bracket of two dynamical quantities $A,B$ which are both integrals  over $M,$ 
 \begin{equation}
 \{ A, B\}_{\star} \doteq \sum_{a} \int _M \mathrm{d}^4 x\,
  \Big[  \Big( \frac{\delta^{\star} A}{\delta q^a(x)} \Big) \, {}^{\;\,\star}_{(x)} \, \Big( \frac{\delta^{\star} B}{\delta p_a(x)} \Big) - \Big( A \rightleftharpoons B \Big) \Big],
  \end{equation}
  where $q^a (x), p_a (x)$ are canonical field components.
  When at least one of $A,B$ contain a $\star$-product, the necessary $\delta^{[\star]}$ variational derivative is generally not defined for non-simple dynamics.
  Therefore $\star$-Poisson brackets are generally $\star$-closed only for simple dynamics $\delta X=0=\delta({\star}).$  This makes canonical quantization of DGR
  challenging; and  even for the simple case the Poisson algebra based dynamics does not describe the classical deformation flow process by which $X_{\mathrm{sc}}$ is attained.\\
 
 The non-associative locking might be circumvented by using the non-standard Lie-derivative mentioned after eqn. (\ref{L1}), which would be a recursive integral twist generally 
 lacking a Hopf algebra structure or Nice Basis.
 But without the Nice Basis Land of Oz properties one cannot define Levi-Civita connections  or even construct gauge (fiber-bundle) theories. The theoretical foundations simply collapse. With apologies
 to Walt Disney and his Wonderful Wizard,  Dorothy might exclaim:``Toto, I don't think we're in Oz anymore!'' \\

 To conclude: For $\delta X\ne0\ne\delta(\star),$ and starting from some initial stationary field configuration $\Psi_0$ for the action, there is generally \emph{no classical field equation} 
 for $\Psi_0$ independent of $\delta\Psi. $
 Hence the non-simple $\delta(\star)\ne 0$ is not viable as a classical \emph{field} theory, even though there still may be stationary action field configurations. 
 Classical DGR must then necessarily be framed within the simple $\delta(\star) = 0$ version discussed 
 earlier, leading to the $\star$-Einstein field equation and $\star$-4-momentum conservation. \\
 
 Now specializing to simple dynamics, the $\star$-analogs of the canonical GR constraints will be derived. To this end, it desirable to change independent variables to make the comparison 
 more transparent. Canonical GR \cite{Rovelli_QG_Chp_2} makes use of the spin connection $\omega^I  _J ,$ a Lorentz Lie algebra $\mathfrak{so}(3,1)$-valued one form. Here the $I,J =0,\ldots ,3$ label 
 the six $\mathfrak{so}(3,1)$ generators as anti-symmetrical pairs, $\omega ^{IJ} = - \omega ^{JI},$ and
 as indices they are raised and lowered by the Minkowski ``metric'' $\eta^{IJ}$ and its inverse matrix. The metrical variables are $\omega$ 
along with the co-tetrads $e^I,$ together comprising a one-form frame bundle. Now going to a base $\star$-manifold $M,$ the $\star$-gauge curvature of $\omega$ is the $\star$-Riemann 2-form,
 \begin{align}
 R[\omega]\, ^{I}_{J} & = \mathrm{d}\,\omega^{I}_{J} + \omega^{I}_{K} \wedge_{\star} \omega^{K}_{J}, \quad\text{ and the action reads}\\
 S[e,\omega] & = \frac{1}{16\pi G} \int_M \epsilon_{IJKL} \Big\{\, \frac{1}{4}\; e^I \wedge_{\star} e^J \wedge_{\star} R[\omega]^{KL} - \frac{\Lambda}{12} \;
 e^I \wedge_{\star} e^J  \wedge_{\star} e^K  \wedge_{\star} e^L \Big\}. \label{ew_action}
 \end{align}
 Here $\Lambda$ is the cosmological constant and $\wedge_{\star}$ is the deformed exterior product between $\star$-forms.  
 The action (\ref{ew_action}) is manifestly $\star$-invariant. Next, introduce the the self-dual
projector $P^{i}_{\;jk} =(1/2) \epsilon^{i}_{\;jk}$, $P^{i}_{\;0j} = - P^{i}_{\;j0} =(i/2) \delta^{i}_{\;j}$ where $i,j,k =1,2,3.$ and extract the self-dual part of $\omega$  as
\begin{equation}
A^i \doteq P^i _{\;IJ}\, \omega^{IJ} = \frac{1}{2} \,\epsilon^i _{\;jk} \,\omega^{jk} + i \,\omega^{0i}.
\end{equation}
The $A^i$ are complex $\mathfrak{so}(3,\mathbb{C})$ algebra-valued $\star$-one-forms. 
The co-tetrads $e^I$ remain real-valued.
Define the $\star$-2-form curvature
 $F^i = \mathrm{d}\, A^i + \epsilon ^{i}_{\;jk} \,A^{j} \wedge_{\star} A^{k},$ then one has the DGR action
\begin{equation}
S[e,A] = \frac{1}{16\pi G} \int_M \Big\{ -i\, P_{jIJ} \,e^I \wedge_{\star} e^J \wedge_{\star} F^j - 
\frac{\Lambda}{12} \, \epsilon_{IJKL}\,e^I \wedge_{\star} e^J  \wedge_{\star} e^K  \wedge_{\star} e^L \Big\}.\label{eA_action}
\end{equation}
To obtain  (\ref{eA_action}) from (\ref{ew_action}), a Nice Basis was used to effect an integration by parts and an integral of a total derivative
was discarded, yielding a $\star$-invariant result. \\

Consider the $\mathfrak{so}(3,\mathbb{C})$-gauge transformation of the $A^i$ generated by three smooth functions $f^i$ as
\begin{equation}
\delta_{f} A^{i}_{\mu} \doteq \mathrm{D}_{\mu} \, f^i \doteq (\mathrm{d}\, f^i )_{\mu} +\epsilon^{i}_{\;jk}\, A^{j}_{\mu} \star f^{k}.
\end{equation}
Then gauge invariance of $S[e,A]$ implies
\begin{align}
0 \doteq \delta_{f} S[e,A] & = \int_M \mathrm{d}^4 x \, \Big[ \delta_{f} A^{i}_{\mu}(x) \Big]\, {}^{\;\,\star}_{(x)} \,\Big[ \frac{\delta^{\star}\, S[e,A]} {\delta A^{i}_{\mu}(x)} \Big]  \nonumber \\
& = - \int_M \mathrm{d}^4 x \, f^{i}(x)  {}^{\;\,\star}_{(x)} \Big[ \mathrm{D}_{\mu} \, \frac{\delta^{\star}\, S[e,A]} {\delta A^{i}_{\mu}(x)} \Big],
\end{align}
and therefore 
\begin{equation}
 \mathrm{D}_{\mu} \,\Big( \frac{\delta^{\star}\, S[e,A]} {\delta A^{i}_{\mu}(x)} \Big) \doteq \partial_{\mu} \Big( \frac{\delta^{\star}\, S[e,A]} {\delta A^{i}_{\mu}(x)} \Big) + \epsilon^{i\;k}_{\;j} A^{j}_{\mu}\, {}^{\;\,\star}_{(x)} \,
\Big( \frac{\delta^{\star}\, S[e,A]} {\delta A^{k}_{\mu}(x)} \Big) = 0. \label{Gauss}
 \end{equation}
 This is the $\star$-analog of the Gauss ($SO(3,\mathbb{C})$-gauge invariance) constraint in canonical GR. \\
 
 Next consider a transformation of $S[e,A]$ under the $\star 4$-diffs generated by four smooth functions $f^{\rho}$ as
\begin{equation}
\delta_{f} A^{i}_{\mu} = f^{\rho} \star \partial_{\rho} A^{i}_{\mu} + A^{i}_{\rho} \star \partial_{\mu} f^{\rho}.
\end{equation}
Specialize to the case where $\mathcal{L}_{X} f^{\rho} = 0$ for the single $\star$-timelike twist generator $X.$ In a Nice Basis (indices wear carets) so that  $\mathcal{L}_{e_{\hat{\mu}}} $ is $\star$-Leibniz, choosing 
$X=e_{\hat{0}},$ and using the Gauss constraint (\ref{Gauss}) one obtains
\begin{equation}
\Big( \frac{\delta^{\star}\, S[e,A]} {\delta A^{i}_{\hat\mu}(x)} \Big) \, {}^{\;\,\star}_{(x)} \,  F^{i}_{\hat{\mu}\hat{\rho}} (x) = 0, \label{4Diff}
\end{equation}
and by the usual argument the carets may now be erased. This \emph{single expression} of $\star$-4-diff invariance replaces the ADM $3+1$ decomposition of standard 4-diffs on undeformed manifolds, 
which read \cite{Rovelli_QG_constaints}
\begin{align}
\Big( \frac{\delta S}{\delta A^{i}_{a} (\vec{\tau})} \Big) \, F^{i}_{ab} (\vec{\tau})= & 0 \quad\text{3-diff invariance, and} \label{3Diff} \\
\Big( \frac{\delta S}{\delta A^{i}_{a} (\vec{\tau})} \Big)\Big( \frac{\delta S}{\delta A^{j}_{b} (\vec{\tau})} \Big) F^{ij}_{ab} (\vec{\tau}) = & 0\quad\text{Wheeler-deWitt.} \label{WdW}
\end{align}
Here $\vec\tau=(\tau_1, \tau_2,\tau_3)$ coordinatizes a 3-slice, the indices $a,b$ live in the 3-slice, and $F^{ij}_{ab}=\epsilon ^{ij}_{\;\;k} F^{k}_{ab}.$
The ADM $3+1$ decomposition fails on $\star$-manifolds because the GM-matter produced timelike $X$ non-locality will not permit data (information) to be separated into 3-slices and 
information normal to those slices. Such a decoupling is not an issue for the ultra-local standard product. This is another example of how non-locality affects the action of $\star$-4-diffs: 
There is no ``$\star$-Wheeler-deWitt equation!'' \\

This disruption of ADM $3+1$ 4-diff decomposition  may be seen from an alternative approach as well.
In the simple dynamics case, one easily derives
\begin{align}
\Big( \frac{\delta^{\star}\, S[e,A]} {\delta A^{i}_{\mu}(x)} \Big) & = P_{iIJ}\,\Big[ \epsilon^{\mu\nu\rho\sigma} \star e^{J}_{\rho} \star  e^{I}_{\sigma} \star n_{\nu}\Big](x), \quad\text{with} \\
n_{\nu} (x) &\doteq \frac{1}{3!} \,\epsilon_{\nu\alpha\beta\gamma} \star \big[\partial_{\tau_1} x^{\alpha}(\vec{\tau}) \big] \star   \big[\partial_{\tau_2} x^{\beta}(\vec{\tau}) \big] \star 
 \big[\partial_{\tau_3} x^{\gamma}(\vec{\tau}) \big] (x).
 \end{align}
 The $n_{\nu}$ are the components of the normal 1-form to the 3-slice.  By anti-symmetry of $\epsilon$,
 \begin{equation}
 n_{\mu}(x)\, {}^{\,\;\star}_{(x)} \, \Big( \frac{\delta^{\star}\, S[e,A]} {\delta A^{i}_{\mu}(x)} \Big) =0. \label{Star_Slice}
 \end{equation}
 The $\star$ in this expression is 4-dimensional, being derived from a GM matter current. As a consequence, the dependence of $S[e,A]$ on $A^{i}_{\mu}(x)$ is \emph{not}
 only from its values on the 3-slice because $\star$ will gather data off the slice when $X$ is timelike. This is unlike the what happens in standard GR \cite{Rovelli_QG_3slice} where the dependence of 
 $S[e,A]$ is \emph{only} through its restriction to the slice; i.e., only through $A^i _a (\vec{\tau}) = (\partial_{a} x^{\mu}(\vec{\tau}) ) \cdot A^i _{\mu}(x(\vec{\tau})).$ That is what (\ref{Star_Slice})
means when $\star$ reverts to the dot-product. This simplification 
 permits the arena of canonical GR to be just the values of fields on a 3-surface, and it fails for DGR. In particular, the canonical gravitational degrees of freedom for DGR are no longer
  the 3-dimensional $A^i _a (\vec{\tau}),$  with the 3-dimensional densitized triad as its conjugate momentum. A related fact is that  there are no well-defined matter generated $3$-dimensional
  twist generator(s) that live only on spatial slices.\\
  
  It may be hoped that the non-canonical deformation flow process (\ref{XSC}) and (\ref{X_Iteration}), by which $X_{\mathrm{sc}}$ is achieved for fixed field configuration $\Psi$ in simple dynamics, 
  could be amenable to deeper understanding 
  by applying the fixed point  theorems mathematicians have developed for metric and Banach spaces. Such theorems include those by Banach, Tychonoff, (non-linear) Krein-Rutman \cite{NL_Krein_Rutman}, 
  and Caristi.
  Regrettably, these powerful statements are all obstructed here by one or more of the following facts: (a) While $\star$-timelike vectors at a point do form a cone $C$ ($\lambda C\subseteq C$ for $\lambda >0$);
  unlike dot-product timelike vectors, they do not comprise a convex set because $C+C \not\subset C.$ (b) $\star$-composite vector fields may not form a complete metric or Banach space. (c) The 
  $X_{\mathrm{sc}}: X^{(n)} \to X^{(n+1)}$ iterative mapping is generally neither (1) a contraction, requiring $d(X_{\mathrm{sc}}(X_1), X_{\mathrm{sc}}(X_2) \,) < d(X_1 , X_2 )$ for some metric 
  $d(X, Y)$ on vector fields; nor (2) $1$-homogeneous, meaning $X_{\mathrm{sc}}(\lambda X) =\lambda X_{\mathrm{sc}}(X),$ for $ \lambda >0.$ As a pedestrian alternative to these elegant theorems, 
  one could use the DDT expansion (IKT within the non-recursive gradient approximation) for $\star$ in $X_{\mathrm{sc}},$ and truncate to a small/finite number of terms. This possesses
  the small expansion parameter $\varepsilon \sim \xi_c ^2\, \|p(\mathrm{GM})\|/L_c$ and convergence of deformation flow is plausible but unproven for $\varepsilon \ll 1.$ Cosmologically,
   this would cover the history of the Universe since well before BBN. But in the ACE when $\xi_c \sim L_c \sim \|p(\mathrm{GM}) \|^{-1},$  $\varepsilon$  is of order unity. During and shortly following that epoch,
   deformation flow might have been non-convergent or wandering, with ill-defined $X_{\mathrm{sc}}.$ This would be somewhat similar to dissipation in classical dynamical systems, such as turbulent flow.
   But here the iterative deformation ``flow''  is not evolving in time. While such speculation might be interesting, it lies beyond the scope of classical DGR because it occurs in the regime where quantum corrections
   to $X$ cannot be safely neglected. \\
   
   \section{Cosmological applications of DGR}
   
   Before taking up $\star$-deformed cosmological models, it is worth clarifying how the classical generator field $X$ is calculated. At the quantum particle level, the $\star$-wave equations for non-GM matter 
   treat $X$ as a classical external field produced by some ensemble of GM matter particles/fields. When considering the the $\star$-wave equations describing the twist producing GM fields themselves,
   $\langle \hat{X}\rangle_{\Omega}$ is an expectation value taken in the GM's state $\Omega.$ 
   This roughly corresponds to a mean field or random phase approximation analogous to the Hartree approximation in atomic physics. 
   Going to the opposite regime where the non-twist length scales are all much longer than the average GM inter-particle  spacing $\zeta_{pp},$
then $\langle \hat{X} \rangle(x) =\mathrm{Tr}\, (\rho(x)\hat {X}  )$ is a smoothly varying local ensemble average over the slowly varying GM momentum distribution $\rho(x).$ This corresponds to the cosmological case studied below.
For length scales below $\zeta_{pp}$ but longer than than $\|p(GM)\| ^{-1},$ one can describe $\langle X\rangle$ as a sum of $\delta$-functions localized at the GM particle positions, each weighted by the 
$\langle \hat{X} \rangle_{\Omega}$ that GM particle is carrying. One then takes an ensemble average over the GM trajectories or momentum distribution. We will not be entering this intermediate regime.
This description of different scales is precisely parallel to how electrical currents carried by charged particles are handled and then suitably averaged, depending on the physical problem. 
In the following, the twist generator $X$ will be treated as the expectation $\langle X\rangle(x)$ taken over an ensemble or distribution of GM particles or their trajectories, 
typically one of local thermodynamic equilibrium. This contrasts with averaging any $\star$-equations over scales much longer than $\zeta_{pp},$ as was performed in earlier work \cite{DGRv1.0}. \\

Cosmological models rely on the theory of maximally symmetric submanifolds. Here these are modified to suit $\star$-manifolds. Suppose the whole $\star$-manifold $N$ is $n$-dimensional, while the maximally symmetric submanifolds $M$ are $m$-dimensional. Let the these sub-manifolds be labelled by $n-m$ coordinates $v^a$, and specify points within each sub-manifold by $m$ coordinates $u^i.$ 
Maximally symmetric means the sub-manifolds having constant $v^a$ are such that the metric tensor of the entire $\star$-manifold is invariant under the group of transformations (isometries)
\begin{align}
u^i & \to u'^i = u^i + \epsilon\,\xi^i (u,v), \\
v^a & \to v'^a = v^a,
\end{align}
with $\epsilon \ll 1,$ a real constant,  $m(m+1)/2$ linearly independent Killing vector fields $\xi^i,$ and $\xi^a (u,v)=0.$ That is, the  $m(m+1)/2$ fields $\xi^i$ are not subject to any linear relations
 with $u$-independent coefficients. The derivation of the following theorem \cite{WGC_Max_Symm_Thm} is valid in a Nice Basis: \\
 
 It is always possible to choose the $u$-coordinates so that the line element of the entire manifold $N$ reads 
 \begin{equation}
 \mathrm{d}s^2 \doteq g_{\mu\nu} \, \mathrm{d}x^{\mu}\,\mathrm{d}x^{\nu} = g_{ab}\, \mathrm{d}v^a \, \mathrm{d}v^b  + f(v)\, \star \, \tilde{g}_{ij} (u)\, \mathrm{d}u^i \, \mathrm{d}u^j ,
\end{equation}
with $\tilde{g}_{ij}(u)$ the metric of an $m$-dimensional maximally symmetric sub-manifold. \\

For example,  to apply this theorem to the 4-dimensional $\star$-FLRW case, the $m=3$ spatial sub -manifolds are maximally symmetric (homogeneous and isotropic for a set of 
fundamental observers), and one sets $v=t,$ $f(v) \doteq a^2 (t) \doteq \tilde{a}^{\star 2}(t)$ to obtain
\begin{equation}
 \mathrm{d}s^2 = -\, \mathrm{d}t^2  + a^2 (t) \star \tilde{g}^{ij}(u) \mathrm{d}u^i \, \mathrm{d}u^j . \label{FLRW_Metric}
\end{equation}
For $N_X=1$  with $\langle X\rangle =X^0 \partial_t ,$  the only $\star$ becomes a dot-product.
This choice of generator arises from a cosmological ensemble average over the $GM$ particles' 
momentum distribution $\rho$ and assumes it preserves the FLRW symmetries. 
Specifically, $\langle X\rangle$ originates from currents of GM particles moving along their timelike trajectories. 
One expects the spatial components of this ensemble average to vanish over cosmological scales so that FLRW homogeneity and isotropy hold.
That is, there may be local places where there are net GM spatial currents, but to zeroth order these are expected not to  survive cosmological (spatial) averaging.
A non-zero averaged GM spatial current component would violate isotropy.
There may be effects of non-zero spatial components of $\langle X\rangle$ on fluctuations about the $0$-order FLRW model, but those lie beyond the 
dynamics of the $0$-order  model itself that is the concern here. Choosing $N_X=1$ is purely a mathematical simplification.
This permits using the FLRW line element (\ref{FLRW_Metric}) in the $\star$-Einstein equation to derive $\star$-Friedmann equations. \\
 
Prior to moving on to those equations, we show how cosmology, and BBN in particular, can be used to constrain models for the twist generator $X.$
When GM matter was introduced \cite{DGRv1.0} two models of $X$ as a gauge currents $J$ were proposed, as expressed in eqns. (\ref{XDPhiN}) and (\ref{XFCN}).
 It was demonstrated there that  $X$ could only be comprised of a $U(1)$
 current since it had to be gauge invariant. Here we consider the $X=J$ proposal further. Since a gauge current must be conserved, we have $X^{\mu}_{\;\;\stackrel{\star}{,} \mu} =0.$
 In a Nice Basis this reads
 \begin{equation}
 |\hat{g}| ^{\star-1/2} \star \Big( \hat{X} \triangleright \big[ |\hat{g}| ^{\star 1/2}  \star X^{\hat{0}} \big] \Big)= 0,  \label{X_Conserv}
 \end{equation}
 where for a $X^j =0$ in a $\star$-FLRW cosmological model we have the Nice Basis $e_{\hat{j}} = \partial_j $ and $e_{\hat{0}} = X = X^0 \,\partial_0 = \partial _{\hat{0}}.$
 Noting that $|\hat{g}| = |g| \star (e_{\star})^{\star2}$ where $e_{\star} = \mathrm{det}_{\star} \, e^{\mu}_{\hat {a}};$ $e^{\mu}_{\hat{a}}$ being the coordinate components of the Nice Basis
 $e_{\hat{a}}:$ $e^{\mu}_{\hat{a}} =\mathrm{diag} \,(X^0 ,1,1,1).$ So $|\hat{g}| = |g| \star (X^0) ^{\star 2}.$ Then from (\ref{X_Conserv}) 
 \begin{equation}
 \partial_0 \big[ \,|g|^{\star1/2} \star X^0\, \big] =0, \quad\text{and thus} 
 \end{equation}
 $X^0 = B |g|^{\star -1/2}$ for $B$ a real constant.  Using (\ref{FLRW_Metric}) one reads off $|g|=\tilde{a}^{\star 6},$ and finally obtains
 \begin{equation}
 X^0 = B\, \tilde{a}^{\star-3}.  \label{X_Conserv2}
 \end{equation}
 If the present value of the non-locality scale is $\xi_c(a=1) \doteq (10^{-33}\,\mathrm{m}) (\xi_c(1)/10^2\, L_P),$ and assuming one can approximate $\tilde{a}^{\star-3} \simeq a^{-3}$ after neutrino decoupling 
 when $T\simeq 1\,\mathrm{MeV}$ at $a\simeq 2\cdot10^{-10},$ one infers the non-locality scale at $\nu$-decoupling: $X\propto \xi_c (\nu\;\mathrm {decoupling}) \simeq (125.\,\mu\mathrm{m} ) 
 (\xi_c(1)/10^2\, L_P).$ Such a long non-locality scale
 would significantly alter BBN involving nucleons with Compton wavelengths $\sim 10^{-15}\,\mathrm{m}.$ To circumvent this problem, either GM is not gauged at all 
 and its current is not necessarily conserved; or GM still carries $U(1)_{GM}$ charge but $X$ is not simply $J_{U(1)}.$ As examples of the latter possibility consider the model expressions
 \begin{align}
 X^{\mu}_{K\phi} &  = \binom {\mathrm{Re}} {\mathrm{Im}} \Big\langle\sum_{l} Q_{l} \, g_{\mu\nu} \star \phi_{l}^{\dagger} \star (\mathrm{D}_{\nu} \phi)_{l} \Big\rangle \star \Big\langle \sum _{l} \phi_{l}^\dagger \star\phi_{l}\Big\rangle^{\star-1} 
\quad\text{ for scalar GM, and} \label{XK_Scalar} \\
 X^{\mu}_{K\psi} &  = \binom {\mathrm{Re}} {\mathrm{Im}} \Big\langle \sum_{l} Q_{l}\, \bar{\psi_{l}} \star \gamma^{I}e_{I}^{\mu}
\star\psi_{l}\Big\rangle\star\Big\langle\sum_{l}\bar{\psi_{l}}\star\psi_{l}\Big\rangle^{\star -1}\quad\text{for fermionic GM, }\label{XK_Fermi}
\end{align}
which modify the original $X=J$ expressions (\ref{XDPhiN}) and (\ref{XFCN}) by the last factor of $\star$-inverse GM number density. The subscript $K$ for Katyusha designates the model's name,
after a popular Russian wartime song. These $X_K$ are still both gauge singlets because the denominators are singlets. Since $X_K$ is no longer a $U(1)$-gauge current, Katyusha pays no heed to  
$X^{\mu}_{\;\;\stackrel{\star}{,} \mu} =0,$ and she successfully evades the BBN constraint. In particular, the denominators in (\ref{XK_Scalar}) and (\ref{XK_Fermi}) vary roughly as $\sim a^{-3}$ 
after GM becomes non-relativistic. This should occur when $T_{GM} \lesssim M_{GM} \sim $TeV scales at about $a\gtrsim 10^{-16}.$ Thus by $a(\mathrm{BBN}) \sim 10 ^{-10}$ the denominator's
$a^{-3}$ variation partially cancels the numerator's $a$ dependence, and continues to do so through the present epoch. 
For scalar GM, one may crudely estimate $X^0 \sim |p(\phi)^0 | \sim T \sim a^{-1},$ so that at BBN,
$\xi_c (\mathrm{BBN}) \simeq (10^{-20}\, \mathrm{m}) (\xi_c(1)/10^5 \,L_P )$ is far shorter than $10^{-15}\, \mathrm{m}$ characterizing nuclear processes then. For fermionic GM, $X_{K\psi}^0$ is 
independent of the scale factor $a,$ and $\xi_c(\mathrm{BBN}) \simeq \xi_c(1).$ In both fermion and scalar cases
 neither BBN nor the subsequent cosmological processes leading to the Cosmic Microwave Background on the surface of last scattering at $a\simeq 9.1\cdot 10^{-4}$ are affected by the twist.\\
 
 By applying simple dimensional analysis to the DDT eqn. (\ref{DDT}) along the lines performed in  \cite{DGRv1.0},   one  estimates the non-locality scale for the Katyusha models as
 \begin{align}
 \big(\xi_c(a=1)/L_P \big)_{K\phi} & \simeq (1.37\cdot 10^3 ) (M_{\lambda}/3\cdot 10^6 \,\mathrm{TeV})^{-2} (M_{\phi}/\mathrm{TeV})^{1},\quad\text{and} \label{Xi_K_Scalar}\\
  \big(\xi_c(a=1)/L_P \big)_{K\psi} & \simeq (1.37\cdot 10^3 ) (M_{\lambda}/9\cdot 10^{12}\, \mathrm{TeV})^{-1},\label{Xi_K_Fermion}
  \end{align}
  where $M_{\lambda}$ is the energy scale of the DDT's dimensionful coupling constant $\lambda.$ Interestingly, the fermionic Katyusha model is independent of $M_{\psi},$ producing 
  $M_{\lambda} \sim \xi_c^{-1},$ a natural scaling. 
 Both the new Katyusha and the previous $X=J$  models are invariant under any combination of gauge transformations, as can be established by appealing to the 
 Hermiticity of the gauge generator matrices. Specifically, GM matter may be SM $SU(2)$-weakly interacting without disrupting the gauge invariance of $X$ or its dependence on $J(U(1)_{GM}).$
 In the earlier work \cite{DGRv1.0} the expressions corresponding to eqns. (\ref{Xi_K_Scalar}) and (\ref{Xi_K_Fermion}) were used to establish that  for the $X=J$ models GM matter could not be any SM
 field. That argument no longer holds for the $X_K$ models because of their different $M_{\mathrm{GM}}$ dependencies of $\xi_c,$ and Katyusha-GM could be either the proton or electron. 
 (Twist producing matter must be stable.) However, that choice is \emph{not} necessary, and Katyusha-GM might instead be furnished by  $U(1)_{\mathrm{GM}}$ gauge interacting particles, 
 which could also be SM $SU(2)$-weakly interacting. Such GM fields provide a viable DM candidate, as was analyzed prviously \cite{DGRv1.0}. So while the Katyusha twist models can account for DM, 
 that is not the unique choice of fields. \\
 
 As a possible alternative to these Katyusha models one might consider GM to be a vector (Proca) field $V^{\mu},$ and set $X_V ^{\mu}= V^{\mu},$ or its real/imaginary part.
 This model is necessarily non-renormalizable. However it does have the virtue that $X_V$ as a fundamental field contains no interior $\star$-products, and then there are no issues about constructing the
 self-consistent $X_{\mathrm{sc}}.$ Similarly no metrical variables enter $X_V$, so upon varying them there is no change in $\star(X_V ),$  and there will be $\star$-field equations for the geometrical
 degrees of freedom as well as for all the SM matter fields. But there is generally no $\star$-field equation for $V^{\mu}$ itself unless once again one imposes simple dynamics $\delta(\star) =0.$ 
 Suppose one does that, then the standard Minkowski spacetime analysis of such a vector field \cite{W_QFT_I_p_320} gives $\partial_{\lambda} V^{\lambda} = m^{-2}\,\partial_{\lambda}\,J^{\lambda},$ 
 where $J^{\lambda}$ is a current coupling as $J_{\lambda} V^{\lambda}$ in the Lagrangian that is produced by other fields. $J^{\lambda}$ might come from $SU(2)$ or $U(1)$ interactions with SM particles, 
 but $X=V$ requires $V^{\mu}$ to be gauge invariant (singlet) to keep $X$ invariant, making $V^{\mu}$ unable to interact with SM fields. 
 Instead setting $J^{\lambda}=0$ leads to the BBN issues discussed above. Namely, then in curved and deformed spacetime 
 $V^{\lambda}_{\;\;\stackrel{\star}{,} \lambda} =0.$ So $X=V$ in a spatially flat $\star$-FLRW cosmological model leads to $\xi_c(a) \sim X^0 (a) \sim \tilde{a}^{\star-3}\sim a^{-3},$ 
 disrupting BBN for plausible values for $\xi_c(a=1).$ Hence to be viable $J\ne 0$ would have to come from an auxilliary set of of non-twist producing GM fields that interact gaugelessly with $V^{\mu}.$
 This complication renders this vector model less appealing than the Katyusha models that interact by dark $U(1)$-GM gauges (dark photons). \\
 
A more exotic alternative possibility is that the twist arises from a condensate. The twist then acquires the DDT form $\mathcal{F} = \exp [\,\theta^{\mu\nu}\, \mathcal{L}_{\mu}\otimes \mathcal{L}_{\nu} ],$
with  $\theta^{\mu\nu} =\langle \,\mathrm{Vac}\, | \hat{\theta}^{\mu\nu} |\, \mathrm{Vac}\, \rangle \ne0.$ But for this to be Lorentz invariant requires $\theta^{\mu\nu} \propto \delta^{\mu\nu}.$ 
This leads to causality issues because it has no vanishing eigenvalues.\cite{DGRv1.0} \\

What are the $\star$-Friedmann equations? Write the $\star$-Einstein equation within a $\star$-FLRW cosmological model in (comoving) cosmic coordinates 
(still following Weinberg' sign conventions \cite{WGC}) as
\begin{align}
 R_{\mu\nu}  & =-8\pi G\, S_{\mu\nu} - \Lambda\, g_{\mu\nu}, \quad \text{with} \label{Friedm_A}\\
 S_{\mu\nu} & \doteq T_{\mu\nu} - \frac{1}{2} \,g_{\mu\nu} \star T^{\lambda}_{\;\lambda} = \frac{1}{2} (\rho - p) \star g_{\mu\nu} + (\rho + p)\star U_{\mu} \star U_{\nu}, 
 \end{align}
 being a prefect fluid source term with the cosmic fluid $4$-velocity in comoving coordinates $U_{\mu} =\delta^{0}_{\mu}.$ $p$ is pressure, and $\rho$ the fluid's energy $3$-density.
 Recall the line element reads
 \begin{equation}
 (\mathrm{d}s)^2 = - (\mathrm{d} t)^2 + a^2(t) \cdot\tilde{g}_{jk}\, \mathrm{d}x^j \otimes \mathrm{d}x^k,
 \end{equation}
 with $a^2(t) = a(t)\cdot a(t) \doteq (\tilde{a} \star \tilde{a})(t),$ and the the $3$-metric $\tilde{g}_{jk}$ is $t= x^0$ independent.
 Assume $N_X =1$ with $X=X^0\, \partial_0,$ coming from an FLRW-symmetric ensemble average over the GM momenta distribution. Straightforward computation leads to
 ($^{\cdot} \doteq \partial_{t}$):
\begin{align}
R_{0jkl} & =0 \\
R_{j0k0} & =\tilde{g}_{jk} \cdot \big[ \dot{a}^2 + a\cdot\dot{a} - (a^2)^{\star -1} \star ( a\cdot\dot{a} ) ^{\star 2} \big] \\
R_{jkjl} & = a^2\cdot \tilde{R}_{jkjl} - ( a\cdot\dot{a} ) ^{\star 2} \cdot \big[ \tilde{g}_{jj}\cdot \tilde{g}_{kl} - \tilde{g}_{lj}\cdot \tilde{g}_{kj} \big]\quad\text{(no j-sum)},
\end{align}
 $\tilde{R}$ is the standard Riemann tensor of the $3$-metric $\tilde{g}_{ij}.$ Using $R_{\mu\nu} = g^{\kappa\lambda} \star R_{\kappa\mu\lambda\nu}$ one finds
 \begin{align}
 R_{00} & = 3\, (a^2)^{\star -1} \star \big[  a^2 + (a\cdot \ddot{a}) - (a^2)^{\star -1} \star (a\cdot\dot{a})^{\star 2}\big] \\
 R_{il} & = -2 k\,\tilde{g}_{il} - \tilde{g}_{il}\cdot \big[ (a\cdot\dot{a})^{\star 2} \star (a^2)^{\star -1} + (\dot{a})^2 + a\cdot\ddot{a} \big] \\
 R_{0i} & = 0,
 \end{align}
 and $k$ is a real constant given by $\tilde{R}_{ij} =-2k\,\tilde{g}_{ij},$ which by a choice of spatial coordinates may be reduced to $k=0, \pm 1.$
 The $00$-component of (\ref{Friedm_A}) reads
 \begin{equation}
 3 (a^2)^{\star -1} \star \big[ \dot{a}^2 + (a\cdot\ddot{a}) - (a^2)^{\star -1} \star (a\cdot\dot{a})^{\star 2} \big] = -4\pi G\,(\rho +3p)+\Lambda,    \label{Friedm_B}
 \end{equation}
 and the $il$-component as
 \begin{equation}
 -2k-\big[ (a\cdot\dot{a})^{\star 2} \star (a^2)^{\star-1} + \dot{a}^2 +a\cdot\ddot{a} \big] = -4\pi G\,a^2\star (\rho -p)-a^2 \Lambda .  \label{Friedm_C}
 \end{equation}
 Using (\ref{Friedm_B}) to solve for $a\cdot\ddot{a},$ and substituting in (\ref{Friedm_C}) leads to the $\star$-Friedmann equation
 \begin{equation}
 (a\star\dot{a})^{\star 2} \star(a^2)^{\star -1} = -k +\frac{8\pi G}{3} \, (a^2)\star \rho + \frac{\Lambda}{3}\, (a^2) \label{Friedm_D}
 \end{equation}
 For $X=0, \star \to \cdot$ this reduces to the familiar $\dot{a}^2 +k =(8\pi G/3) \rho a^2 +(\Lambda/3)a^2.$
 One may re-express (\ref{Friedm_D}) in terms of the Hubble parameter $H(a)$ as 
 \begin{equation}
H^2 \doteq (\dot{a} \cdot a^{-1})^2 =-ka^{-2} +\frac{8\pi G}{3} \, \big(\rho\star (a^2)^{\star 2} \big) \cdot \big(a^{-4}\big) + \frac{\Lambda}{3}\, \big(a^2\big)^{\star 2} \cdot \big(a^{-4} \big). \label{Friedm_E} 
\end{equation}
Defining $\rho_c(t) \doteq (3/8\pi G) H^2,$ $\Omega \doteq \rho/\rho_c,$ and $\Omega_{\Lambda} \doteq \Lambda/3 H^2$ with $\rho_{\mathrm{tot}} \doteq \rho + (\Lambda/8\pi G),$ this 
may be recast as
\begin{equation}
\Big| \frac{\rho_{\mathrm{tot}} \star(a^2)^{\star 2} } {\rho_c \cdot a^4 } -1 \Big| = \frac{|k|}{(a\cdot H)^2} .
\end{equation}
So if $(aH)$ decreases with increasing $t,$ as in the standard radiation and matter dominated epochs, $\Omega_{\mathrm{tot}} \doteq \rho_{\mathrm{tot}} /\rho_c$ 
seems finely tuned to unity in the early Universe, and the $\star$-product by itself does not offer a solution for the cosmological flatness problem. \\
 
 To understand the $\star$-evolution of the scale factor $a$ from eqn. (\ref{Friedm_E}) better requires more information about the thermal equilibrium $\rho(a).$
 In an Nice Basis ($\;\hat{{}}\;$) use the perfect fluid form for the energy-momentum tensor 
 $T_{\hat{\mu}\hat{\nu} } =(p+\rho) \star U_{\hat{\mu}} \star U_{\hat{\nu}} + p\star g_{\hat{\mu}\hat{\nu} }$  in the conservation law $T^{\hat{\mu}\hat{\nu}} 
_{ \;\;\;\stackrel{\star}{,}\hat{\nu}} =0,$ together with the $\star$-FLRW form for $g_{\hat{\mu}\hat{\nu}}$ to obtain 
\begin{equation}
\tilde{a}^{\star 3} \star \partial_{\hat{0}} p = \partial_{\hat{0}} \big[ \tilde{a}^{\star 3} \star (p + \rho) \big]. \label{Friedm_E_Cons}
\end{equation}
Within a $\star$-FLRW model the thermal equilibrium entropy $S$, $3$-volume $V$, and temperature $T$, together with $\rho(T)$ and $p(T)$ are only dependent on $x^0$ or $x^{\hat{0}},$ resulting in
\begin{equation}
T\star \mathrm{d}S(V,T) = \mathrm{d}\big[ \rho(T)\star V \big] +p(T)\star \mathrm{d}V. \label{Freidmn_E1B}
\end{equation}
Thus $(\partial S/\partial V)_T = T^{\star -1} \star (\rho +p)$ and $(\partial S/\partial T)_V = V \star T^{\star -1} \star (\mathrm{d}^{\star}p/\mathrm{d}T)$.
Here we have defined $\mathrm{d}^{\star} p/\mathrm{d} T \doteq f$ iff $\mathrm{d} p = f \star \mathrm{d} T.$
 Integrability $(\partial^2 S/\partial \,V\partial T ) 
 = (\partial^2 S/\partial \,T\partial V )$ requires 
 \begin{equation}
 \big(\partial/\partial T\big)_V \big[ T^{\star-1} \star (\rho +p)\big ] = \big(\partial/\partial V \big)_T \big[ V\star T^{\star-1} \star(\mathrm{d}^{\star} \rho/\mathrm{d}T) \big]. \label{Friedmn_Integ2}
 \end{equation}
 By the $\star$-chain rule $\mathrm{d}^{\star} p/\mathrm{d}T =(\mathrm{d}^{\star} T/\mathrm{d}x^{\hat{0}}) ^{\star-1} \star (\mathrm{d} ^{\star} p/\mathrm{d}x^{\hat{0}})$ and also using the $\star$-Leibniz property of a Nice Basis
 the RHS of (\ref{Friedmn_Integ2}) becomes
 \begin{align}
\big (\partial/\partial V\big)_T  \big[ V \star T^{\star -1}\star(\mathrm{d}^{\star} \rho/\mathrm{d}T) \big] & = 
\big(\partial V/\partial x^{\hat{0}}\big)^{\star -1} \star \big(\partial / \partial x^{\hat{0}}  \big)_T \big [ V \star T^{\star -1}\star(\mathrm{d}^{\star} \rho/\mathrm{d}T) \big] \nonumber\\
& = (T^{\star -1})\star (\mathrm{d}^{\star} \rho/\mathrm{d}T).
\end{align}
Likewise the LHS of (\ref{Friedmn_Integ2}) may be written as
\begin{align}
\big( \partial/\partial T\big)_V \big[ T^{\star -1} \star (p+\rho ) \big] & =
\big(\partial T/\partial x^{\hat{0}} \big)^{\star -1} \star \big( \partial / \partial x^{\hat{0}} \big)_V \big[T^{\star -1} \star (p+\rho) \big] \nonumber\\
& = - T^{\star -2} \star (p+\rho) +T^{\star -1} \star\big[ \mathrm{d}^{\star} (p+\rho)/\mathrm{d}T \big].
\end{align}
Hence $ -T^{\star -2}\star(p+\rho)  + (T^{\star -1}) \star (\mathrm{d}^{\star} p/\mathrm{d}T) =0$ or
\begin{equation}
\frac{\mathrm{d}^{\star} p} {\mathrm{d} T} = T^{\star -1} \star (p+\rho). \label{Friedmn_E3}
\end{equation}
$\star$-multiplying this by $\tilde{a}^{\star 3} \star (\partial T/\partial x^{\hat{0}}),$ using eqn. (\ref{Friedm_E_Cons}), and recalling for a Nice Basis with $N_X =1$ and $e_0 = X$
that $(\mathrm{d}^{\star} /\mathrm{d}x^{\hat{0}}) =
(\mathrm{d} /\mathrm{d}x^{\hat{0}})$  leads to 
\begin{equation}
T\star \partial_{\hat{0}} \big[ \tilde{a}^{\star 3} \star (p+\rho) \big] = (\partial T/\partial x^{\hat{0}}) \star \tilde{a}^{\star 3} \star (p+\rho). \label{Friedmn_E4}
\end{equation}
Hence, again by the $\star$-Leibniz property of a  Nice Basis,
\begin{align}
\frac{\partial}{\partial x^{\hat{0}} } \big[ \tilde{a}^{\star 3} \star T^{\star -1} \star (p+\rho) \big] & = T^{\star -1} \star \frac{\partial}{\partial x^{\hat{0}} } \big[ \tilde{a}^{\star 3} \star (p+\rho) \big] -T^{\star-2} 
\star  \frac{\partial T}{\partial x^{\hat{0}} } \star \tilde{a}^{\star 3} \star (p+\rho) \nonumber \\
& = T^{\star -1} \star \frac{\partial}{\partial x^{\hat{0}} } \big[ \tilde{a}^{\star 3} \star (p+\rho) \big] -T^{\star -1} \star \frac{\partial}{\partial x^{\hat{0}} } \big[ \tilde{a}^{\star 3} \star (p+\rho) \big] \nonumber \\
& =0. 
\end{align}
Here  (\ref{Friedmn_E4}) was used to go from the first to second lines. Now changing independent variables from $x^{\hat{0}}$ to $x^{0},$ one finds
\begin{equation}
 \tilde{a}^{\star 3} \star T^{\star -1} \star (p+\rho) = \text{constant}. \label{Friedmn_E5}
 \end{equation}
 Then using (\ref{Friedmn_E3}) in (\ref{Freidmn_E1B}) yields 
 \begin{align}
\mathrm{d} S(V,T) & = T^{\star -1} \star \mathrm{d} \big[ (\rho(T)+p(T)) \star V \big] -T^{\star -2} \star V \star (p+\rho) \star (\mathrm{d} T) \nonumber \\
& = \mathrm{d} \big[ (p+\rho) \star V \star T^{\star -1} \big].
\end{align}
So $S(V,T) = (p+\rho)\star V\star T^{\star -1} + \mathrm{constant}, $ and (\ref{Friedmn_E5}) states the constancy of entropy within a $3$-volume $V \propto \tilde{a}^{\star 3}.$
When all the particles in thermal equilibrium are ultra-relativistic in the comoving frame (during the radiation epoch) $p=\rho/3,$ and eqn. (\ref{Friedmn_E3}) reads 
\begin{align}
\mathrm{d}\rho &= 4 \rho \star T^{\star -1} \star (\mathrm{d}T),\quad\text{or} \nonumber \\
0 & = \mathrm{d} \big[ \ln_{\star} \rho - 4 \ln_{\star} T \big]  = \mathrm{d} \ln_{\star} \big( \rho \star T^{\star -4} \big),
\end{align}
 And finally  $\rho \propto T^{\star 4}$ during the radiation epoch, a result that one might have intuitively conjectured.
 Now using this result in eqn. (\ref{Friedmn_E5}), one finds $\tilde{a}^{\star 3} \star T^{\star 3} =\mathrm{constant},$ or $T \propto \tilde{a}^{\star -1},$ 
 and $p,\rho\propto \tilde{a}^{\star -4}$ in that epoch.
 Returning to (\ref{Friedm_E}) and setting $k=0=\Lambda$ for simplicity leads to
 \begin{align}
 H^2 = (\dot{a}\cdot a^{-1})^2 & =(8\pi G/3) (\rho \star \tilde{a}^{\star 4}) \cdot(a^{-4}) \nonumber \\
 \dot{a}^2 & = (8\pi G/3) \rho_0 a^{-2} \nonumber \\
 \dot{a} & = \mathbb{A}\, a^{-1},
 \end{align}
for constants $\rho_0, \mathbb{A}.$ This has the standard flat radiation solution $a \propto t^{1/2}.$ The twist does not seem to evade the $t=0$ singularity classically.
For the general $\Lambda =0$ case one obtains
\begin{equation}
\dot{a}^2 = -k  +\frac{8\pi G\rho_0}{3} \, a^{-2},
\end{equation}
the same as the standard dot-product FLRW model.
This absence of twist effects in the evolution of $a$ is remarkable. This especially true for the the IKT, which would require $a(t<0)$ data to evaluate $\star$-products involving $a$ or $\tilde{a}$
at early times when $t\lesssim \xi_c.$
It is not possible na\"ively to  restrict the range of the $t_1, t_2$ integrals interior to the IKT to enforce $t_1, t_2 \ge0$ because then the $\star$-product would no longer be unital in that epoch.
However, in all models for matter produced twists, the GM modes are frozen at sufficiently small $a$ when $H(a) \gtrsim \| p_{\mathrm{GM}} \|.$ For instance, for $X_{K\phi} \propto 
\partial_0 \phi ,$ $\| X\| \to 0$ as $a\to 0^{+},$ so $\mathrm{d}\sigma^{\mu} /\mathrm{d}t =X^{\mu} \to 0,$ and the IKT cannot ``reach back'' to $t\le 0.$ When the GM currents or modes freeze, $\star$
becomes like the dot-product.\\
 
Could a DGR as classical non-local theory resolve the cosmological horizon problem without appeal to inflation? This might have occurred in the very early Universe when
the size of the currently entire observable Universe $d_{\mathrm{obs}}(a)$ was on the order of or smaller than $\xi_c(a).$  During that epoch non-local micro-causality violating processes
could have homogenized what later became the (nearly) cosmologically homogeneous Universe we observe today.  This epoch is referred to as the acausal epoch (ACE). 
We now examine this scenario starting with fermionic  Katyusha model. The size of the current observable Universe is $\approx 29.0$ GLy, and it  it scales linearly with $a.$
Since $\xi_c (a)$ scales with $X,$ which for this model is $a$-independent,  the condition $d_{\mathrm{obs}}(a^{*}) \simeq \xi_c(a^{*})$ reduces to
\begin{equation}
a^{*}_{K\psi} \simeq (5.8\cdot 10^{-60}) (\xi_c(1)/100\,L_P)
\end{equation}
with $\xi_c(1)$ the current value of the non-locality scale. However if one can extrapolate the radiation epoch relation $T(a) \sim (1 \mathrm{MeV})(a/10^{-10})^{-1}$ to such tiny $a,$
then $T(a^* ) \sim 10^{27}\, E_P,$ placing this well into the Planck epoch for plausible values of $\xi_c(1),$ and then the fermionic model cannot explain the cosmological horizon problem. \\

There is initially more hope for the scalar Katyusha model where $\xi_c(a) \sim X(a) \sim |p^0(\mathrm{GM})| \sim T \sim a^{-1}$ for relativistic GM when 
$a\lesssim 3\cdot 10^{-16}(M_{\mathrm{GM}}/\mathrm{TeV})^{-1}.$ As a result
\begin{equation}
a^{*}_{K\phi} \simeq (2.4\cdot 10^{-30}) (\xi_c(1)/100\,L_P)^{1/2}
\end{equation}
and $T(a^{*}) \sim (3.4\cdot 10^{-3}\,E_P) (\xi_c(1)/100\,L_P)^{-1/2}.$ One also finds 
\begin{equation}
H(a^*)/|p^0_{\mathrm{GM}} (a^*)| \simeq (0.102)\,(\xi_c(1)/100\,L_P)^{-1/2},
\end{equation}
and then the scalar GM modes are not cosmologically frozen at $a^* .$
 Similarly one can use the expression for $X_{K\phi}$ to estimate when the ACE ended by  $\xi_c(a_{\mathrm{ACE}}) \simeq H^{-1} (a_{\mathrm{ACE}}),$ and then arrive at
 \begin{align}
 a_{\mathrm{ACE}} & \simeq (5.9\cdot 10^{-21}) (\xi_c(1)/100\,L_P)^{1/3} \quad\text{with} \\
 T(a_{\mathrm{ACE}}) & \simeq (1.7\cdot10^4 \,\mathrm{TeV}) (\xi_c(1)/100\,L_P)^{-1/3}.
 \end{align}
 From this one can see that $a^*$ occurred very early in the ACE,  At $a=a^*$,  the non-locality scale was quite long: 
$ \xi_c(a^*) \simeq (6.6\cdot10^{-4}\,\mathrm{m})\,(\xi_c(1)/100\,L_P)^{1/2} \gg  \| p_{\mathrm{GM}} \|^{-1}$
 This signals a $\star$-self-averaging situation for the scalar field $\phi$ itself. To see this, derive the $\star$-scalar wave equation in a Nice Basis using simple dynamics as
 \begin{align}
 P[\phi] &\doteq  |\hat{g}|^{\star-1/2} \star \partial_{\hat{\mu}} \big[ g^{\hat{\mu}\hat{\nu}} \star  |\hat{g}|^{\star1/2} \star \partial_{\hat{\nu}}\, \phi \big] -m^2\,\phi = 0, \quad\text{or}\label{Star_scalar1}\\
 P[\phi] & = \big[ g^{\hat{\mu}\hat{\nu}} \star \partial_{\hat{\nu}} \, \phi \big] _{\stackrel{\star}{,} \hat{\mu}}  -m^2 \, \phi = 0. \label{Star_scalar2}
 \end{align}
 As usual, (\ref{Star_scalar2}) can now be extended beyond the Nice Basis. Applying this to the $\star$-FLRW model,  specializing to $k=0=\Lambda,$ with $N_X =1$ and
 $X=X^0\, \partial_0$,  the $\star$-scalar wave equation reduces to
 \begin{equation}
 P[\phi] = -m^2\,\phi + \tilde{a}^{\star-2} \star (\nabla^2\,\phi) - (\partial_0)^2\, \phi -3\,(\partial_0 \tilde{a}) \star (\tilde{a}^{\star-1}) \star (\partial_0 \,\phi).\label{Star_scalar3}
 \end{equation}
 Thus when
 $ \xi_c(a^*)  \gtrsim  \| p_{\mathrm{GM}} \|^{-1}$ the scalar GM field will be self-averaged by $\star$ to become diffusive or frozen over $\xi_c(a^*),$ similar to a cosmologically 
 diffusive or frozen mode over $H^{-1}(a)$ when $H(a) \gtrsim \| p_{\mathrm{GM}} \| .$ Consider what occurs as one ``travels backwards'' in cosmic time: For the Katyusha $\phi$ model, 
 this means  that as $a$ decreased from  where GM became  relativistic, 
 $\xi_c(a) \sim a^{-1}$ increased and $\| p_{\mathrm{GM}} \|^{-1} \sim T^{-1}\sim a$ decreased. But this trend must have terminated when $\xi_c(a_{\mathrm{pk}}) \simeq 
 \| p_{\mathrm{GM}} \|^{-1} (a_{\mathrm{pk}}).$
Then as $a$ decreased further, $\xi_c(a)$ stayed pinned to  $\| p_{\mathrm{GM}} \|^{-1} (a),$ giving $\xi_c(a)\sim X^0(a)\sim a$ for $a\lesssim a_{\mathrm{pk}}.$ 
This limit of  $\|X(a)\|\to 0$ as $a\to 0^+$ also prevents the IKT from trying to access data from $t<0.$
One can roughly estimate 
$a_{\mathrm{pk}}$ to find that the currently observable Universe then had size $d_{\mathrm{obs}} ( a_{\mathrm{pk}} ) \simeq (2.5\cdot 10^{11}\,\mathrm{m})(\xi_c(1)/100\,L_P)^{1/2} \gg
\xi_c(a_{\mathrm{pk}} ).$ Thus the observable Universe never was completely inside $\xi_c(a),$ and  classical DGR acausal effects could not have homogenized it.\\

Hence neither the scalar nor fermionic Katyusha models can resolve the cosmological horizon problem.
The ACE ends just where classical DGR is becoming unreliable due to non-negligible quantum corrections to $X,$ and its further investigation lies beyond the scope of this work.\\

Because the non-locality over the length scale $\sim\xi_c$ introduced by the $\star$-product induces a breakdown of the cluster decomposition principle that underlies the 
ladder operator formalism of local quantum field theory, 
it would be interesting to calculate the inverse (Green's) operator $\Delta$ for the $\star$-scalar wave equations (\ref{Star_scalar1}) - (\ref{Star_scalar3}).  $\Delta$ contains information on how
the cluster decomposition principle and micro-causality are recovered at scales longer than $\xi_c$ since it governs the field commutator $[\hat{\phi}(x), \hat{\phi}(y)]=i\,\hat{1}\, \Delta(x,y).$ 
Specifically, in standard quantum field theory $\Delta(x,y)$ vanishes if $x,y$ are space-like separated. But for DGR this is disrupted on the length scale $\xi_c.$
Regrettably attempts to perform this calculation
have been unsuccessful.
Just as in the dot-product case, the nonlinear coupling of $\tilde{a}$ or other metrical variables to $\phi$ dynamics complicates the analysis.  
Neglect that coupling, and $\star$-effects disappear.
One can introduce a $\star$-analog of the Fourier transform 
based at some point $x$  for $N_X =1$ as
\begin{equation}
(\hat{f})_{x}(q) \doteq \exp\, \big[\beta(q)\big] \int _{- \infty}^{\infty} (\mathrm{d}\hat{t}/2\pi) \exp\big[ -i q\, \hat{t}\, \big] f[\sigma_X (x,\hat{t} ].
\end{equation}
 Just as in the dot-product case, using the IKT this reduces the wave equations to integral equations in $q$-space, replacing $\star$ throughout by the \emph{standard} convolution of the transforms.
 However the nonlinearity then simply becomes  an equally intractable non-locality in $q$-space, obstructing computation of integral kernels such as $\Delta(x,y).$ 
 The disruption of the ladder operator algebra is discussed  further in section 9 below. \\
 
It is possible, however, to make \emph{qualitative} statements about the consequences of  ladder algebra disruption and failure of micro-causality on the length scale $\sim \xi_c.$
In particular, anti-particles are formally introduced using the ladder algebra to construct quantum field theory on Minkowski spacetime so that it is both Lorentz invariant and compatible
 with the cluster decomposition principle/micro-causality. Could DGR have consequences for anti-matter? Consider the following argument from Weinberg's classic text \cite{Weinb_Antip}:
 The quantum probability of a particle of mass $m$ reaching an event $x_2$ if it starts at event $x_1,$ even if $x_1, x_2$ are space-like separated in Minkowski spacetime, is non-negligible as long as 
 \begin{equation}
 0< (\vec{x}_1-\vec{x}_2)^2 -(x_1^0 -x_2^0)^2 \lesssim (\hbar /m)^2\doteq\big(\Lambda_C(m)\big)^2,
 \end{equation}
 where $\Lambda_C$ is the particle's Compton wavelength.
 So if one observer sees the particle emitted at $x_1$ and absorbed at $x_2,$ then another (boosted) observer paradoxically detects the particle absorbed a $x_2$ at time $t_2$ before the emission time 
 $t_1$ at $x_1.$ To resolve this paradox, the second observer sees a particle emitted at $x_2$ and absorbed at $x_1,$ and that particle is necessarily different from the seen by the first observer, 
 generally with oppositely signed quantum numbers, namely the anti-particle. 
 Now consider the situation in DGR, replacing $\Lambda_C$ in the above equation with $\xi_c$ when $x_1$ and $x_2$ lie on the same TSM. In that case the emission and absorption events are no longer
 physically distinct \emph{events.} That is, for timelike generator $X,$ it is not \emph{physically} meaningful to speak of separable emission and absorption \emph{events} even if the coordinates $x_1$ and $x_2$ 
 are \emph{mathematically} separable \emph{points.} This was the essence of  the $\star\delta$-function action discussed at the end of section 3 via the IKT.
 So at length scales $\lesssim\xi_c$ there is no more ``paradox'' to resolve, and the physical necessity for anti-particles disappears for energies $\gtrsim \xi_c^{-1}.$ This lack of anti-particles at such energies
 could have played a role in the very early Universe and offers a qualitative explanation for the presently observed matter-antimatter asymmetry. 
 There simply was no significant amount of anti-matter in that very early epoch. This phenomenology is robust with respect to the choice of generator model for the matter-based twist, however it would require $N_{X}=2$
 so that the TSMs contain spacelike separated points.
 \\
 
 Two decades ago the causal dynamical triangularization approach to quantum gravity discovered the reduction of spacetime spectral dimension \cite{Causal_Dyn_Triang} from 4 to 2 on the Planck scale $L_P.$
 The idea of this phenomenon  is that a diffusion-like process
 describing the probability of return to a point on the real spacetime manifold $\mathcal{M}$ provides a dynamical dimension not necessarily equal to the standard (Lebesgue covering) dimension
  $n =\dim \mathcal{M}.$
 So consider the $n$-dimensional diffusion equation $\partial_s \,\phi(s,x) = -D \,\Box^{(n)} (x)\, \phi(s,x),$ where $s$ is some fictitious (nonphysical) ``time,'' $\Box^{(n)}\, \phi \doteq P[ \phi ]$ is 
 the $n$-dimensional massless scalar wave
 operator on $\mathcal{M}$ such as  eqn. (\ref{Star_scalar2}), and
 $D$ is a positive real (diffusion) constant.  The initial  condition at $s=0$ is  $ \phi (0,x) = \delta ^{(n)}(x).$ On standard $n$-dimensional Euclidean space, 
 $\Box^{(n)}$ is the $n$-dimensional Laplacian, and one has the well-known the Gaussian solution
 \begin{align}
 \phi(s,x) & = \Big( \frac{1}{2\pi \sigma(s)^2 } \Big)^{n/2} \exp\big[ -x^2/2\sigma(s)^2 \big],\quad\text{and} \label{SDR_Gaussian} \\
 \sigma(s)^2 & = 2 D s.
 \end{align}
 If $\phi (s,x)$ is interpreted as the probability of return to $x$ by ``time'' $s,$ then the spectral dimension $n_s$ can be read off from (\ref{SDR_Gaussian}) as
 \begin{equation}
 n_s = -2 \lim_{s\to +\infty} \ln \phi(s,0)/\ln s\,.
 \end{equation}
 Since the initial simulations demonstrated spectral dimension reduction, researchers  have provided phenomenological insights into this intriguing behavior \cite{LM_Phenom}.
 In particular, interpretations involving the large momentum scaling of the scalar wave operator $P[\phi ]$  have been presented. So far all these descriptions are quantum in nature.
For classical DGR there is a deformation of the usual scalar wave operator due to its internal $\star$-products. According to the IKT representation of $\star$ there are special 
 $N_X $-dimensional submanifolds, namely the TSM. Could the the averaging \emph{only} on the TSM lead to classical spectral dimension reduction on the length scale $\sim \xi_c \gg L_P ?$\\
 
 The massless $\star$-scalar wave operator of interest here was derived earlier as (\ref{Star_scalar2}), namely
 \begin{equation}
 P[\phi]  = \big[ g^{\mu\nu} \star \partial_{\nu} \, \phi \big] _{\stackrel{\star}{,} \mu}
 \end{equation}
 in any coordinate system. One is interested in the scaling of this operator at large $4$-momentum $k^{\mu}$ of $\phi.$ Since the $\star$-inverse metric $g^{\mu\nu}$ is roughly like the reciprocal
 of  an area $A,$ one can crudely conceive of this as a scaling of $1/A(\ell)$ with length $\ell .$ In the case of quantum gravity, one makes use of the loop quantum approach result that 
 the area operator has minimum eigenvalues of order $L_P^2 ,$ and sets $A(\ell) =\ell^2 + L_P^2.$ Putting $k\simeq 1/\ell,$ the inverse-area scaling is described by
 \begin{equation}
 \mathcal{F}(k) =\big[ A(1/k, L_P ) \big]^{-1} / \big [ A(1/k_0, L_P ) \big]^{-1} 
 \end{equation}
 where $k_0 \ll L_P$ is some ``small'' fiducial momentum where one fixes $\mathcal{F}(k_0 )= 1.$
 Then
 \begin{equation}
 \mathcal{F}(k) =\frac{ (k^{-2} + L_P^2 )^{-1}}  { (k_0^{-2} + L_P^2 )^{-1}} = \frac{k^2}{1 + (L_P \,k)^2} \cdot \frac{1 + (L_P\, k_0)^2 } {k_0^2}.
 \end{equation}
 One finds $\mathcal{F}(k) \simeq (k/k_0)^2$ for $k_0 \ll k \ll1/L_P$ and $\mathcal{F} \simeq (k_0 L_P)^{-2}$ for $k \gg 1/L_P \gg k_0.$
 The $k$-independent saturation of $\mathcal{F}(k)$ for $k\gtrsim L_P$ is the hallmark of spectral dimension reduction at the Planck scale.\\
 
 Here we apply this phenomenological approach to DGR. $\xi_c$ takes on the role of $L_P$ because as the initial condition for $\star$-diffusion $\delta^{\star}_x$
 acts approximately as a Gaussian of extent $\xi_c$ only on the TSM$(x).$ Also approximate $P_{\phi}(k) \simeq - g^{\mu\nu}\star (k_{\mu} k_{\nu} \phi).$ 
 Separating  $P_{\phi}(k) $ into its action on the $N_X$-dimensional TSM and the orthogonal $(n-N_X)$ directions, one may write
 \begin{align}
 \mathcal{F}(k) & \simeq \frac{1}{n}\, \Big[ \;\sum_{a=1}^{N_X} \frac{k_a^2}{1+(\xi_c \,k_{\parallel})^2}  \frac{1+(\xi_c \,k_0)^2} {k_0^2}  +  \sum_{a=N_X +1}^{n} \frac {k_a^2} {k_0^2}\; \Big] \\
 & \simeq \frac{1}{n}\,\Big[ \,\frac{N_X \,k_{\parallel}^2}{1+(\xi_c\, k_{\parallel})^2} \frac {\big(1+(\xi_c \,k_0)^2\big)} {k_0^2}  + \frac {(n-N_X)\,k_{\perp}^2}{k_0^2} \,\Big]. \label{SDR_F}
 \end{align}
 Here $k_{\parallel}$ refers to the projection of $k$ onto the TSM, $k_{\perp}$ is the off-TSM part.
  The overall $1/n$ factor maintains $\mathcal{F}(k_0) =1$ for $k_0 \ll 1/\xi_c.$ For $k_0 \ll |k_a| \ll\xi_c^{-1},$ $\mathcal{F}(k) \simeq k^2/k_0^2 .$
  For $\phi$ propagation on the TSM one drops the second term on the RHS of (\ref{SDR_F}) and
  \begin{equation}
  \mathcal{F}_{\parallel} (k) \simeq \frac{1}{n} \,\Big[\, \frac{N_X \,k_{\parallel}^2}{1+(\xi_c\, k_{\parallel})^2} \,\frac{1}{k_0^2} \,\Big] \to \frac{N_X}{n(\xi_c \,k_0)^2} \quad\text{for}\quad
  k_{\parallel} \gg \xi_c^{-1}.
  \end{equation}
  This large $k$ behavior saturates, signaling dimensional reduction.
  Similarly off TSM one drops the first term to obtain
  \begin{equation}
  \mathcal{F}_{\perp}(k) \simeq\frac{n-N_X}{n}\, \Big(\frac{k_{\perp}}{k_0}\Big)^2,
  \end{equation}
  which never saturates, and so is not dimensionally reduced.
  For $k\simeq k_{\perp} \simeq k_{\parallel}$ one gets
  \begin{equation}
  \mathcal{F}(k\simeq k_{\perp} \simeq k_{\parallel} ) \simeq \frac{n-N_X}{n} \,\Big(\frac{k}{k_0}\Big)^2 \quad\text{for}\quad  k\gg\xi_c^{-1}
  \end{equation}
  a non-saturating limit and thus not dimensionally reduced. 
  Hence one has found classical anisotropic scalar diffusion, but not an overall reduction of spectral dimension as in the quantum case. Only diffusion on the TSM exhibits
  spectral dimension reduction. This micro-anisotropy of spacetime will be explored further in the next section. \\
 
 \section{Experimental Tests of DGR}
 
 Due to the non-locality of DGR both the celebrated Spin-Statistics Theorem (SST) and CPT-symmetry from standard quantum field theory are expected to be violated. 
 As a result of the breakdown of the spin-statistics relationship, one also anticipates deviations from the Pauli Exclusion Principle, for which there are stringent experimental bounds
 \cite{Addazi}. One way to approach SST breakdown is to study the parameter $\alpha_{ij}$ governing the exchange of identical particles $i,j$ in a multi-particle state $\Phi.$
 Here we assume that spacetime is adiabatic in the Birrell-Davies sense \cite{Birrell_Davies}: The characteristic frequencies $\omega$ of particles (matter modes) are much larger than the rates
  of change of metrical variables, such as $H(a).$ It is then permissible to speak of particles with their associated quantum numbers $q$ that comprise the state $\Phi,$ which is taken
  to be a ray of vectors in some Hilbert space. \\
  
  Suppose one exchanges two identical particles $i$ and $j$ in $\Phi,$ so $\Phi \to \alpha_{ij} \,\Phi$ for complex $\alpha_{ij}.$  If $\Phi$ is normalized to unity, then $|\alpha_{i,j}| = 1.$
  Upon repeating the exchange, one returns to the original state, hence $\alpha_{ij} \,\alpha_{ji} =1.$ In standard quantum field theory $\alpha_{ij}$ is taken to be a function of a \emph{single}
  particle property, namely to depend solely on the species $n=n_i =n_j$ of particles $i$ and $j.$ This way $\alpha_{ij}(n) = \alpha_{ji}(n),$ and then $(\alpha_{ij})^2 =1,$ yielding $\alpha_{ij} =\pm 1$
  and the familiar Bosonic and Fermionic quantum statistics. In a non-local theory with characteristic length $\xi_c,$ $\alpha_{ij}$ may depend on other particles within a distance $\sim\xi_c$ 
  of $i,j$, as ``influential spectators'' $s.$ So $\alpha_{ij}$ depends not only on the quantum numbers $q_i , q_j$ of $i,j,$ respectively, but possibly also on $q_s;$ i.e., on all the quantum numbers 
  specifying $\Phi.$ $\alpha$ is no longer simply a single particle property of the exchanged particles. However if one arranges all those quantum numbers into a list, then the ordering of that list
  of numbers on which $\alpha$ depends is not expected to be physically relevant. That is, although $\alpha$ itself has well-known physical effects, the value of $\alpha$ is not 
  anticipated to depend on how theorists choose to list the quantum numbers on which $\alpha$ depends. Making that assumption, $\alpha_{ij}$ must depend on a symmetrized 
  combination of all the quantum numbers specifying $\Phi.$ In particular, $\alpha_{ij} = \alpha_{ji},$ once again arriving at $\alpha_{ij}= \pm 1.$ Here one is also assuming that $\alpha$ 
  does not depend on any paths taken to assemble $\Phi,$ such as anyons which can only occur in two spatial dimensions. 
  As Weinberg \cite{W_QFT_I_exchange} has pointed out, $\alpha_{ij}$ cannot depend only on the spins $\sigma_i ,\sigma_{j}, \sigma_{s}$ because then $\alpha$
  would form a $1$-dimensional representation of the rotation group, and there are none. Nevertheless, there is still room for novel effects. 
  Even with symmetrized $\alpha$ in $3+1$-dimensions, one could have $\alpha_{ij} = f((p_i -p_j )^2),$ some function of the $4$-momentum exchange.  But since $\alpha_{ij} =\pm 1,$
  any changes in $\alpha_{ij}$ must be \emph{discontinuous.}  This discontinuous change would be expected to occur in a non-local theory at characteristic
  $4$-momenta $\sim\xi_c^{-1}.$  The possibility that $\alpha$ is no longer simply a single particle species dependent constant causes multi-particle states to possess a potential 
  ``split-personality disorder'' as far as their statistics are concerned. A state comprised of two identical spin-$0$ particles may behave quite decorously (Dr. Jekyll) as 2 bosons below some threshold
   $4$-momentum exchange, only to act as  a nefarious fermion pair (Mr. Hyde) above threshold! Within DGR $\xi_c^{-1}$ is expected to lie in the range $(10^{-6} -10^{-2}) E_P,$  and experiments at
   smaller characteristic energies will be censored from witnessing the personality changes. Specifically, this would apply to the tests of the Pauli Exclusion Principle performed to date, since their
   typical energies are on the atomic to nuclear scales,  at most $10-100$ MeV, many orders of magnitude beneath any expected non-locality induced threshold(s).\\
   
   The familiar ladder operator $\hat{a}(q), \hat{a}^{\dagger}(q)$ algebra for a particle with quantum numbers $q$ is derived from the $\alpha_{ij}$ taken as a single particle species 
   property \cite{W_QFT_I_ladder_op}. For instance one constructs $N$-particle states from the particle vacuum state $\Phi_0$ as $\hat{a}^{\dagger}(q_1) \cdots \hat{a}^{\dagger}(q_N) \Phi_0.$ In the standard 
   ladder algebra, where $\hat{a}(q)$ is defined as the adjoint of $\hat{a}^{\dagger}(q),$ if $q, q_1, \ldots , q_N$ are all sub-threshold bosons (upper sign choice) or all fermions (lower sign) one finds
   \begin{equation}
   \hat{a}(q)\; \Phi_{q_1  \ldots  q_N } = \sum_{r=1}^{N} \,(\pm 1)^{r+1} \,\delta(q-q_r )\, \Phi_{q_1 \ldots q_{r-1} q_{r+1} \ldots q_N}.
   \end{equation}
   Above threshold this becomes disrupted as soon as $\alpha_{ij}$ becomes dependent on the influential spectators $s\ne i,j$ in an $N$-particle state, since then $\alpha_{ij}$ can be $N$-dependent. 
   Consequently above threshold $\hat{a}(q)\,\Phi_0 \ne 0$ becomes possible along with 
   \begin{equation}
   [\hat{a}(q'), \hat{a}^{\dagger}(q)] _{\mp}\; \Phi_{q_1 \ldots q_N} \ne \delta(q'-q)\, \Phi_{q_1 \ldots q_N}
   \end{equation}
    and
   $[\hat{a}(q'), \hat{a}(q)] _{\mp} \ne 0.$ The  physical process underlying this is that the ladder algebra encodes the cluster decomposition principle. This is violated by DGR since $[\hat{\psi}(x),
   \hat{\psi}(y)]_{\mp} \ne 0$ for space-like separated $x,y$ within a proper distance $\sim\xi_c$ of each each other. The ladder algebra is an inappropriate foundation for quantum field theory above 
   threshold. In that strange land one must bid adieu to the familiar Feynman diagram technology, loops, and so on.\\
   
   While spin-statistics, the Pauli Exclusion Principle, and the ladder operator algebra arise from particle exchange effects on multi-particle \emph{states,} CPT symmetry is a symmetry of a
    local Lorentz invariant \emph{action}  under transformation of the \emph{fields} as $\psi\to (CPT)\, \psi.$ In a non-local theory CPT symmetry will be broken. But unlike $\alpha_{ij}=\pm 1$ or 
    $|\alpha_{ij}|=1$ for particle exchange, there are no corresponding  thresholding restrictions for CPT violations. In fact it is easy to show that 
    \begin{align}
    (CPT) \big\langle \hat{X}^{\mu}(x) \big\rangle (CPT)^{-1} & \ne \big\langle \hat{X}^{\mu}(x) \big\rangle \quad\text{and} \\
    (CPT) (\star_{X}) (CPT)^{-1}& \ne (\star_{X}).
    \end{align}
    What is expected size of the CPT violations? In an experiment with some typical $4$-momentum $p,$ the first order $\star$-interactions relative to the standard dot-product ones is
    \begin{equation}
    \eta(\mathrm{CPT}) \simeq \frac{\xi_c^2\, \|p\|}{L_c},\label{eta_grav}
    \end{equation}
    with $L_c$ the gravitational curvature radius at the experiment. On the Earth's surface, $L_c\simeq (c^2 R^3/GM)^{1/2} \simeq 10^{11}\,\mathrm{m},$ and then
   $ \eta(CPT)\simeq (5\cdot 10^{-53}) (\xi_c/10^5 L_P)^2 (E(p)/\mathrm{TeV}).$
    Near a black hole with $L_c \simeq 10^4$ m, $\eta(CPT)$ is enhanced by a factor $10^7,$ but even that really does not help the experimental situation much. 
    At the end of the ACE in the very early Universe, $\eta(CPT) \sim 1.$\\
    
    The current experimental bounds on $\eta(CPT)$ are too insensitive to detect such minuscule violations. The Brookhaven measurement of the sidereal variations
    of the muon magnetic moment's $g-2$ found no detectable violations down to $10^{-24}$ GeV, or $\eta(CPT)\lesssim 10^{-23}$ at $100$ MeV energies \cite{Brookhaven}.
    The corresponding DGR estimate for flat spacetime is  of second order in momentum: $\eta\sim( \| p\| \, \xi_c)^2 \lesssim 10^{-28}.$
    At larger energies, experiments bounding the mass difference between the top and anti-top quark
    masses to $67.1$ ppm at $7$ TeV imply $\eta(CPT) \lesssim 6.7\cdot 10^{-5}$ at that energy \cite{Top}. Even if CPT violations are ever detected, they could arise from violations of Lorentz invariance
    or non-locality, so by themselves they would not constitute unambiguous evidence for DGR. \\
    
    One may also consider deviations from maximum particle velocities being bounded by $c$ that could arise from $\star$-wave equations and their
     violation of the cluster decomposition principle. These can only occur in curved spacetimes, coming from the kinetic term in the $\star$-action. 
     This results in the current experimental bounds being at least $24$ orders
     too insensitive to the $\star$-corrections relative to the undeformed contributions, which theoretical ratio is about the same as $\eta(CPT)$ from eqn. (\ref{eta_grav}). \\
     
     Yet another class of possible experimental tests are modern Michelson-Morley type experiments that search for anisotropies in the phase velocity of photons. DGR can produce an effect like this but
     again only in curved spacetimes as the photon propagation direction varies with respect to the relative velocity of the local flow of GM particles. The 2009 experimental bound \cite{MM_09} 
  on this anisotropy is an impressive $\lesssim 10^{-17}.$  On the other hand, the DGR estimate for photon wavelength $\lambda$ on Earth is
  \begin{equation}
  \eta (MM) \simeq \frac{\xi_c^2}{L_c \lambda} \simeq (5.1\cdot 10^{-65}) \,(\xi_c/10^5 L_P)^2 \,(\lambda/500\,\mathrm{nm})^{-1},
  \end{equation}
 still a long way beyond present technology.\\
 
 Hughes-Drever experiments are precision spectroscopic measurements of the anisotropy of space and mass. First performed in 1960-1, they are among the most accurate tests of relativity.\cite{Hughes_Drever}
  The twist might affect nuclear transitions, here between Zeeman levels in an external magnetic field, because the transition frequencies could potentially become anisotropic 
  respect to the twist generator(s). By comparing resonant 
  Larmor frequencies in a ${}^{129}\mathrm{Xe}/{}^{3}\mathrm{He}$ co-magnetometer, Allmendinger et al. \cite{Allmendinger}  determined $\Delta M/M < 6.7\cdot 10 ^{-34}$ for the nucleon. 
  Both these nuclei are spin $1/2,$ so by the Wigner-Eckart theorem they have no electric/magnetic quadrupole or higher moments. 
  A twist produced externally to the nuclei would alter  the nuclear magnetic moments so the ratio of their Larmor frequencies in the same external magnetic field would vary with time as $X$ in the lab frame 
  changed with the Earth's sidereal or annual motions. If the twist arises from GM matter that is the Dark Matter comprising the galactic halo, then
  $(\Delta M/M)_{\mathrm{DGR}} \simeq (\xi_c(\mathrm{meas})/\Lambda_C(N) )^2 \simeq (\xi_c(\mathrm{meas})/10^3 \,L_P)^2 (10^{-32}).$ Here $\Lambda_C(N)$ is the nucleon Compton wavelength, and 
  $\xi_c(\mathrm{meas}) =\xi_c(X)\,(\beta\gamma),$ where the special relativistic factor $(\beta\gamma) \simeq 6.9\cdot 10^{-4}$ enters because $X=X^0\, \partial_0$ in the GM or halo's rest frame 
  while the Earth moves through it along its galactic orbit at $v\simeq 206$ km/sec. Allmendinger et al's 2013 $\Delta M/M$  bound implies $\xi_c(X_{\mathrm{DM}}) /L_P <3.76\cdot 10^5 ,$ 
  consistent with DGR expectations. On the other hand, if stable Standard Model particles such as nucleons or nuclei generate the twist in their own rest frame, then by symmetry that 
  local twist generator vector must lie parallel/anti-parallel to the particle's spin, even during the spin's resonant dynamics. Hence in that case the magnetometer nuclei will perceive no sidereal or annual shifts in the
  ratio of their Larmor precession frequencies, producing a null experimental result, and no bound on $\xi_c(X).$ 
  Nevertheless, future $4-5$ orders of magnitude improvements in Hughes-Drever type experimental precision and accuracy could possibly rule out $X$ coming from 
  non-standard matter by bounding $\xi_c(X_{\mathrm{DM}})/L_P \lesssim \mathscr{O}(1).$ \\
  
  \section{Conclusions and perspective}
  
  So where are we? What has been learned about DGR, what remains obscure? DGR has been outfitted with a curved spacetime compatible integral kernel, whose convergence and equivalence to
  an extended DDT have been clearly delineated. Practical working details of $\star$-coord reparams ($\star4$-diffs) have been laid out and applied to develop firm foundations for classical $\star$-gauge theories 
  ($\star$-Cartan structure and $\star$-Bianchi identity) and $\star$-Einstein field equations for simple dynamics $\delta X=0 =\delta(\star).$ The non-simple case was demonstrated to be non-viable
  as a classical field theory. Photon and graviton ghosts were banished from Minkowski spacetime. $\star$-FLRW cosmological models were studied, and an explicit solution was found in the radiation dominated epoch
  for the spatially flat case without a cosmological constant.
  BBN constraints were applied to constrain
   the expressions for the twist generators $X.$ DGR does provide viable models for dark matter as the twist producing GM matter such as the Katyusha forms for $X.$ In those cases, dark matter 
   would likely be a plasma of dark-$U(1)$ interacting scalars or fermions. DGR can also qualitatively account for the matter-antimatter asymmetry of the observable Universe.
 It further predicts thresholded violations of the Spin-Statistics Theorem and Pauli Exclusion Principle that lead to an unexpected 
   Dr. Jekyll-Mr. Hyde high energy statistical behavior.  Future Hughes-Drever measurements hold out some promise as experimental tests of whether twists are produced by standard matter or not.\\
   
   However for each of these ``successes'' there are corresponding unresolved issues, no-shows, or loud calls of ``out of bounds.'' 
   When applied to cosmology
DGR's $\star$-FLRW models provide no resolution for the horizon or flatness problems.
   There is no $\star$-ADM $3+1$ decomposition of $\star 4$-diffs, and the use of gauge 
   holonomies is restricted to loops. The elegant methods of knot invariants and spin networks, so successful for quantizing canonical GR, are unfortunately useless for DGR. 
   Micro-causality and the cluster decomposition principle are strongly violated 
   on the  $\xi_c(a=1)\simeq 10^{3-5} \,L_P$ length scale,  yet attempts to explicitly calculate how those cornerstones of quantum field theory are recovered at longer lengths were all rebuffed.
   The intriguing non-canonical deformation flow process $X^{(n)}_{\mathrm{sc}} \to  X^{(n+1)}_{\mathrm{sc}}$ also remains a challenging unsolved problem.
   Looking towards future theoretical developments, the method of 
   algebraic quantum field theory has shown some recent progress as being adaptable to $\star$-manifolds with fixed classical $X_{\mathrm{sc}}.$ 
   It is able to enter into the  (Birrell-Davies) non-adiabatic regime, where the particle concept (ladder operator algebra) and micro-causality are not valid, 
   but regrettably it does not address deformation flow. 
   A detailed understanding of the predicted non-locality dominated early cosmological acausal epoch (ACE)  has posted ``No Trespassing!"  writ  large for classical DGR. 
   Finally, aside from the potentially hopeful exceptions of Hughes-Drever and CPT-violation measurements, 
   the possible experimental tests of DGR examined so far lie many orders of magnitude  beyond present technology.\\
   
   To place these virtues and defects of DGR into better perspective, we present a brief discussion within the larger context of non-local modified theories of gravity.
   These kinds of theories have been considered at least since the late 1980's.\cite{Krasnikov}\cite{Tomboulis}  They consist of gravitational actions made from  an infinite set of 
   higher derivative operators acting on curvature terms, and go by the name of infinite derivative gravity (IDG) theories. Theories that just add a finite number of higher order curvature terms like $R^2$ to the 
   Einstein-Hilbert action are necessarily local.
   The most general 4-dimensional, torsion-free, and parity symmetric IDG action around a constant curvature manifold was derived by \cite{Biswas_1}, and reads
   \begin{equation}
   S  = \int \mathrm{d} ^{4}x\, |g|^{1/2} \, \Big[ M_P^{2} \,R + R \,F_{1} (\Box) \,R + R^{\mu\nu} \,F_{2} (\Box) \,R_{\mu\nu}  + C^{\mu\nu\lambda\sigma} \,F_{3} (\Box) \, C_{\mu\nu\lambda\sigma} \, \Big],
   \end{equation}
   with $C$ the Weyl tensor, $F_i (\Box) =\sum _{k=0}^{\infty} f_{ik}(\Box/M^2)$, numerical coefficients $f_{ik}$, $\Box = g^{\mu\nu} \nabla _{\mu} \nabla _{\nu}$
   (d'Alembertian), and UV mass scale $M$.   
   To be ghost free, the $F_i$ must all be exponentials of entire functions.\cite{Buoninfante_etal} Bounds on $M$ were obtained by \cite{Edholm} and \cite{Feng} as $M\gtrsim 10^{-5} \, E_P$, the same order of magnitude
   as $\xi_c^{-1}$ in DGR. In contrast to IDG, theories have also been developed possessing infra-red nonlocalities, partly to understand the cosmic acceleration and/or cosmological constant 
   problems.\cite{Barvinsky} There also have been results exploring infra-red nonlocal covariant actions that include MOND gravity.\cite{Soussa_Woodward} Since DGR is UV nonlocal, 
   we focus on comparing
   it to IDG. Both theories are background independent and respect (local) Lorentz invariance. However, IDG is invariant under standard local diffs, 
   while DGR is invariant under GM matter controlled, nonlocal $\star$-diffs. IDG is UV-complete and  (super) renormalizable/finite \cite{Modesto_IDG};
   but classical DGR is an effective non-renormalizable theory, valid only up to energy
   scales $\sim \xi_c^{-1}$. IDG's milder covariant d'Alembertian-based nonlocality leaves the light cone invariant and preserves the cluster decomposition principle and micro-causality. 
   By contrast, DGR is more disruptive and violates  these on length scales $\lesssim \xi_c$. IDG's nonlocality only affects the gravitational sector,
   while DGR's $\star$ enters all interactions in order for the total action to be $\star$-diff invariant. The standard ADM $3+1$ decomposition is expected to be valid for IDG
    since it has been reported to possess a Hamiltonian formulation.\cite{Talaganis}
    That decomposition does not obtain for DGR. So far IDG has made no statements about dark matter, but DGR provides
    natural models for it. In the large energy limit, IDG displays quantum spectral dimension reduction, whereas DGR cannot reliably attain that quantum regime, but instead shows
    classical anisotropic diffusion effects. Since DGR violates the cluster decomposition principle on short length scale, it can accommodate early universe matter-antimatter asymmetry,
    but IDG is silent about this.  Both IDG and DGR nonlocalities violate the Raychaudhuri null focussing theorem. Nevertheless, IDG still exhibits  spacetime solutions that include singularities.\cite{IDG_exact}
    On the other hand, DGR encounters predictability/renormalization issues before reaching any singularities, and its current version makes no claim of singularity avoidance. 
 Both  IDG and DGR require some early inflationary epoch to resolve the cosmological horizon problem. Taken as a whole, IDG possesses a moderate nonlocality that does not challenge long
 cherished theoretical concepts such as renormalizability.
 Its elegant UV completeness arises because its form of nonlocality efficiently suppresses high energy quantum spacetime fluctuations, and is then more appealing to those who have conviction in renormalizability. 
 However, so far it does not shed much light on observational puzzles.
 By contrast, DGR was explicitly designed to break the cluster decomposition principle on short lengths 
 so micro-causality becomes emergent rather than fundamental.  \cite{Minkowski_Procs17} It can account for some of those puzzles, yet is presently limited to being an effective theory. \\

   ``In every journey, there are as many objectives missed as there are objectives gained.'' --Arnold Toynbee, 1960 \\
   
   \section{Acknowledgements}
   
   The author would like to thank Patrizia Vitale for bringing the work of her research group about commutative deformations on Minkowski spacetime to his attention.  Although at first glance those integral kernel methods
   seemed inappropriate to gravitational physics, in the end they indeed pointed out a way forward. \\

 \end{document}